\documentclass[namedreferences]{solarphysics}
\pdfoutput=1

\ifx\pdfoutput\undefined
\usepackage{graphicx}
\else
\usepackage[pdfauthor={J.~Threlfall},citecolor=blue,linkcolor=blue, linktocpage=true,colorlinks=true]{hyperref}
\usepackage{epstopdf}
\fi 

\usepackage[optionalrh]{spr-sola-addons} 
\usepackage{graphicx}        
\usepackage{color}           
\usepackage[usenames,dvipsnames,svgnames,table]{xcolor}
\usepackage{subfig}
\usepackage{cases}
\graphicspath{ {./figs/} }

\DeclareRobustCommand*{\unit}[1]{\def~{\,}\ensuremath{\mathrm{\,#1}}}

\chardef\us=`\_

\begin{document}

\begin{article}
\begin{opening}

\title{Flux Rope Formation Due to Shearing and Zipper Reconnection}

\author[addressref=aff1,corref,email={jwt9@st-andrews.ac.uk},]{\inits{J.}\fnm{J.}~\lnm{Threlfall}}
\author[addressref=aff1,email={awh@st-andrews.ac.uk}]{\inits{A.~W.}\fnm{A.~W.}~\lnm{Hood}}
\author[addressref=aff1,email={erp@st-andrews.ac.uk}]{\inits{E.~R.}\fnm{E.~R.}~\lnm{Priest}}
\address[id=aff1]{School of Mathematics and Statistics, Mathematical Institute, University of St Andrews, St Andrews, KY169SS, U.K.}

\runningauthor{J. Threlfall \textit{et al.}}
\runningtitle{Flux Rope Formation Due to Zipper Reconnection}

\begin{abstract}
{Zipper reconnection has been proposed as a mechanism for creating most of the twist in the flux tubes that are present {prior to} eruptive flares and coronal mass ejections. We have conducted a first numerical experiment on this new regime of reconnection, where two initially untwisted parallel flux tubes are sheared and reconnected to form a large flux rope. We describe the properties of this experiment, including the linkage of magnetic flux between concentrated flux sources at the base of the simulation, the twist of the newly formed flux rope and the conversion of mutual magnetic helicity in the sheared pre-reconnection state into the self-helicity of the newly formed flux rope.} 
\end{abstract}

\keywords{ Magnetic Reconnection, Observational Signatures;
                  Magnetic Reconnection, Theory;
                  Magnetic fields, Corona;
                  Flares, Relation to Magnetic Field}
\end{opening}

\section{Introduction}\label{sec:intro}
{Eruptive solar flares and coronal mass ejections (CMEs) contain large highly twisted magnetic flux ropes, and a key question is how is the twist created, since the twist in the pre-eruptive state is much smaller than what is observed later in the solar wind in an interplanetary coronal mass ejection (ICME) or magnetic cloud (MC) \citep{paper:Webb2000,paper:Demoulin2008,paper:Vourlidas2014}.

{Prior to the onset of an eruption, the magnetic structure around a prominence is highly sheared and slowly evolves through a series of equilibria. A magnetic flux rope is often present, or formed during the earliest stages of the eruption. Many studies of the build-up to the onset of flares and CMEs have been undertaken, with a flux rope being formed in several ways including flux emergence \citep{paper:ArchontisHood2008}, cancellation of flux \citep{paper:vanBallegooijenMartens1989}, and reconnection at a quasi-separator \citep{paper:Aulanieretal2010} or a separator \citep{paper:LongcopeBeveridge2007}. Our investigation will study the flux rope formation process itself, rather than the eruption onset either due to a loss of equilibrium \citep{paper:PriestForbes1990,paper:LinForbes2000} or an equilibrium instability \citep[\textit{e.g.}][]{book:PriestForbes,book:Priest} often attributed to either the kink instability \citep{paper:HoodPriest1979,paper:FanGibson2003,paper:Toroketal2004} or the torus instability \citep{paper:TorokKliem2005,paper:KliemTorok2006}, which controls most of the dynamic and explosive behaviour witnessed during a flare or CME. 

It is generally accepted that one effect of the eruption process is to drive reconnection in the region below the erupting flux rope. Such} underlying reconnection may either create a new flux rope or increase the flux and twist of a pre-existing weakly-twisted rope \citep{paper:Gibsonetal2004,paper:Gibsonetal2006}. The erupting prominence is invariably seen to be more highly twisted than before it erupted \citep[\textit{e.g.}][]{paper:Mackayetal2010,paper:MackayYeates2012}. Furthermore, three-dimensional (3D) reconnection naturally tends to create twist \citep{paper:BergerField1984, paper:HornigPriest2003, paper:Priestetal2016}.
  
The characteristics of flux ropes observed in interplanetary space have also been compared with the properties of their source regions at the Sun. In such cases, the poloidal flux has been found to be roughly equal to the total magnetic flux that is reconnected at the Sun \citep{paper:Qiuetal2007}, suggesting that flux ropes in the interplanetary medium are mainly created at the Sun during the initial phase of a CME by magnetic reconnection. An interesting feature of solar flares and CMEs \citep[described by, \textit{e.g.}][]{paper:Fletcheretal2004} is that they have two distinct phases in which the nature of reconnection changes \citep{paper:Yangetal2009, paper:Qiuetal2010}. During the rise phase, a flare brightens in H$\alpha$, at one point on each flare ribbon, and there is a single flare loop connecting the two points in extreme ultra-violet (EUV) and hard X-rays. These chromospheric brightenings spread rapidly along the polarity inversion line (PIL) to form the flare ribbons and at the same time a whole arcade of flare loops is formed whose footpoints lie in the ribbons. The ribbons and arcade are created by what \citet{paper:PriestLongcope2017} call ``zipper reconnection". Then the main phase of the flare is created by ``main-phase reconnection" and is characterised by the outwards motion of the ribbons in a direction perpendicular to the PIL and the rise of the flare arcade. Zipper reconnection explains the initial motion of flare brightenings parallel to the PIL. It builds up magnetic twist and flux in the core of the erupting flux rope during the rise phase. Then, during the main phase, reconnection adds more twist and flux to the central core of the flux rope. Concepts of magnetic helicity conservation and the initial configuration itself are used to quantify the build-up of both twist and magnetic flux in the erupting flux rope \citep{paper:PriestLongcope2017}.

Zipper reconnection has features in common with ``flux cancellation'' \citep{paper:Martinetal1985,paper:vanBallegooijenMartens1989} and ``tether cutting'' \citep{paper:Mooreetal2001, paper:Amarietal2003a, paper:Amarietal2003b,paper:Aulanieretal2010} eruption models, but two new factors in our analysis are the role of magnetic helicity conservation in calculating the resulting flux rope twist, and the progression of the reconnection along the PIL. Flux cancellation refers to the cancellation of magnetic flux at the photosphere usually driven by photospheric motions towards the PIL and leading to the formation of a flux rope (and subsequently a prominence), whereas zipper reconnection can also be initiated after motions parallel to the PIL, can occur above the photosphere, and is involved in the rise phase of a two-ribbon flare. Tether cutting reconnection was proposed as a means of initiating a flare by reconnecting two sheared flux ropes under a coronal arcade to form a sigmoid, whereas zipper reconnection refers to the reconnection of a series of loops placed along and inclined to the polarity inversion line and is a response to the eruption.

In light of such models, several recent papers are particularly relevant to our work. \citet{paper:Wangetal2017} combine 
solar and interplanetary observational data to give a detailed overview of the dynamic formation of a single magnetic flux rope in the solar corona during a two-ribbon flare. Crucially, these authors derive estimates for how the poloidal and toroidal flux and the field line twist evolve over time. From a modelling perspective, \citet{paper:Inoueetal2018} use cutting-edge techniques, combining non-linear force-free field (NLFFF) extrapolations (based upon magnetogram images of an active region which produced an M6.6-class flare) and magnetohydrodynamic (MHD) simulations of the formation and eruption of a magnetic flux rope due to tether-cutting reconnection. While much focus is placed on the eruptive phase of the flux tube development, these authors also evaluate the build-up of magnetic twist both before and after the tether-cutting reconnection forms the flux rope.}

In the present paper, we set up a numerical experiment to study zipper reconnection and determine whether it can indeed form a large flux rope. In Section~\ref{sec:config} we describe our simple numerical model, where a pair of magnetic bipoles (initially linked by a pair of untwisted magnetic flux tubes) are sheared and undergo magnetic reconnection to form a twisted overlying magnetic flux rope. We study several properties of this resulting configuration and its evolution over time in Section~\ref{sec:results}; these properties include the magnetic flux and connectivity at the base of the model (in Section~\ref{subsec:base}), the formation and evolution of the flux rope (in Section~\ref{subsec:form}) and an analysis of the flux rope twist and self helicity (in Section~\ref{subsec:twist}). We discuss our findings and outline our conclusions and future work in Section~\ref{sec:conc}. 

\section{Computational Setup}\label{sec:config}

\begin{figure}[t]
\centering
 \resizebox{0.99\textwidth}{!}{\includegraphics{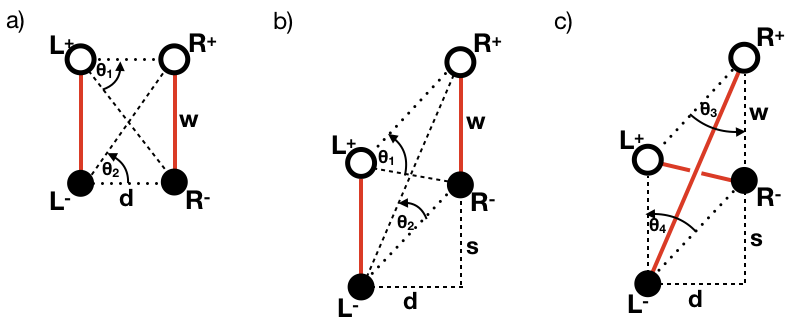}}
  \caption{Evolution of the basic configuration (seen from above) of our experiment, where two flux tubes are sheared perpendicular to the PIL. In (a) a pair of uniform flux tubes connect regions of opposite magnetic polarity. The flux tubes are identified by solid red lines and link the left positive (L$^+$) and negative (L$^-$) or right positive (R$^+$) and negative (R$^-$) regions of strong vertical magnetic field. Positive and negative regions are separated by a distance $w$, while regions of the same polarity are separated by a horizontal distance $d$. In (b) these regions are sheared (by a distance $s$) perpendicular to the polarity inversion line. In (c) new flux ropes form in the newly sheared configuration due to reconnection.}
 \label{fig:zip2}
\end{figure}

We model a single ``zippette'' flux rope formation by starting with two untwisted flux tubes beside each other and allow them to relax to reach a near-potential state. Once the overall field has relaxed, one tube is then moved with respect to the other so
that the whole configuration is skewed, as shown in Figure~\ref{fig:zip2}. Unlike other similar previous experiments, the shear phase alone is responsible for the configuration that evolves once reconnection begins. Previous experiments combine the shear phase with inflow towards the PIL in order to initiate reconnection. Such an inflow is not necessary in our experiment.

The evolution of the magnetic field is followed by solving the MHD equations
\begin{eqnarray}
\frac{\partial \rho}{\partial t} + \nabla \cdot (\rho \bf{v}) &=& 0\; ,\label{eq:continuity}\\
\rho \frac{\partial \bf{v}}{\partial t} + \rho \bf{v} \cdot \nabla {\bf v} &=& - \nabla p + \bf{j} \times \bf{B}\; , \label{eq:motion}\\
\frac{\partial \bf{B}}{\partial t} &=& \nabla \times (\bf{v} \times \bf{B}) + \eta \nabla^2 \bf{B}\; , \label{eq:induction}\\
\frac{\partial p}{\partial t} + {\bf v}\cdot \nabla p &=& - \gamma p \nabla \cdot {\bf v} + \eta j^2\; ,\label{eq:energy}
\end{eqnarray}
where $\rho$ is the plasma density, $p$ the plasma pressure, ${\bf{v}}$ the velocity, $\bf{B}$ the magnetic field, ${\bf{j}} = \nabla \times \bf{B}/\mu$ the current density, $\eta$ the magnetic diffusivity, $\gamma = 5/3$ the ratio of specific heats and $\mu = 4\pi \times 10^{-7}$ is the magnetic permeability. In addition, $\nabla \cdot {\bf B} = 0$.
These are solved using the Lagrangian remap code, ({{Lare3D}}) described in \citet{paper:LareXd2001}. {{Lare3D}} uses dimensionless variables, based on a magnetic field strength, $B_0 = 10$ G, a length, $L = 10^{7}$m, and a mass density, 
$\rho_0 = 1.67 \times 10^{12}\unit{kg\  m}^{-3}$. Thus, the typical Alfv\'en speed is $V_{\rm{A}} = 690$ km s$^{-1}$ and the typical time is $t=14.5$ s. The reference current density is $j_0 = B_0/(\mu L) = 8\times 10^{-5}\unit{A\  m}^{-2}$ 
and the reference magnetic diffusivity is $\eta_0 = 6.9 \times 10^{12}\unit{m}^{2}\unit{s}^{-1}$.

The magnetic diffusivity consists of two parts, namely a background value, $\eta_{b}$, and an anomalous value, $\eta_{\rm{a}}$, that is only non-zero when the current rises above a critical value. 
This allows the formation of strong currents and ensures that the reconnection remains spatially local. In dimensionless variables, 
\begin{equation}
\eta = \left \{
\begin{array}{cc}
\eta_{b}  & |j| < j_{\rm{crit}}\; ,    \\
\eta_{a} + \eta_{b}  &  |j| \ge j_{\rm{crit}}\; ,  \\
\end{array}
\right .
\end{equation}
and $\eta_{b}= 10^{-4}$ and $\eta_{a} = 10^{-3}$.

In the subsequent sections, time units are quoted in Alfv\'en times, $\tau_{\rm{A}} = L/V_{\rm{A}}$, and lengths in terms of $L$. The computational domain considered is: $-8 \le \bar{x} \le 8$,$-8 \le \bar{y} \le 8$ and $0 \le \bar{z} \le20$ (where barred quantities represent dimensionless variables in the numerical domain). 
\begin{figure}[t]
 \centering
   \begin{minipage}[b]{0.47\textwidth}
    \resizebox{\textwidth}{!}{\includegraphics{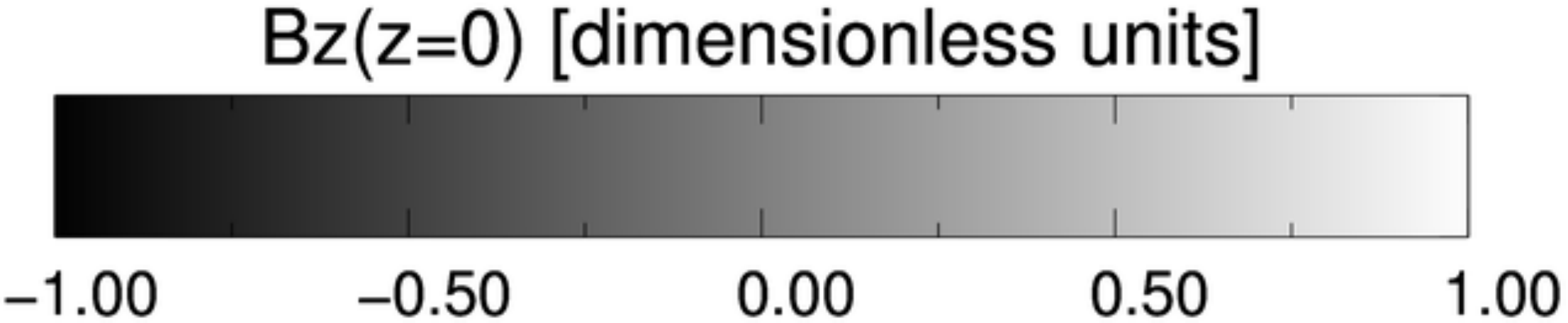}}\\
    \subfloat[Initial connectivity ($B_z\ge0.37$)]{
   \label{subfig:base000}\resizebox{\textwidth}{!}{\includegraphics[clip=true, trim=25 10 35 65]{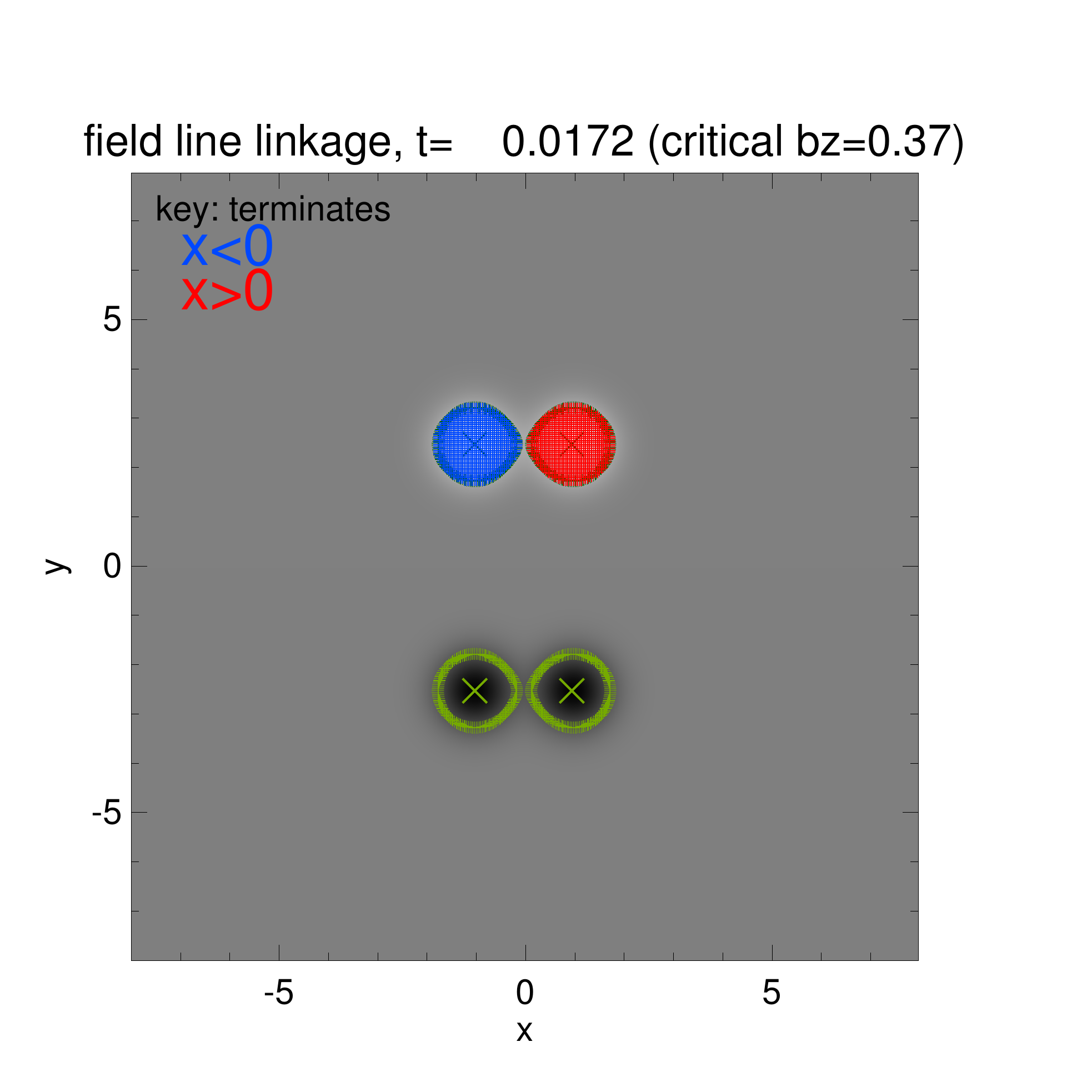}}}
   \end{minipage}
   \begin{minipage}[b]{0.51\textwidth}
     \subfloat[Initial 3D magnetic field configuration]{\label{subfig:cfg000}\resizebox{\textwidth}{!}{\includegraphics{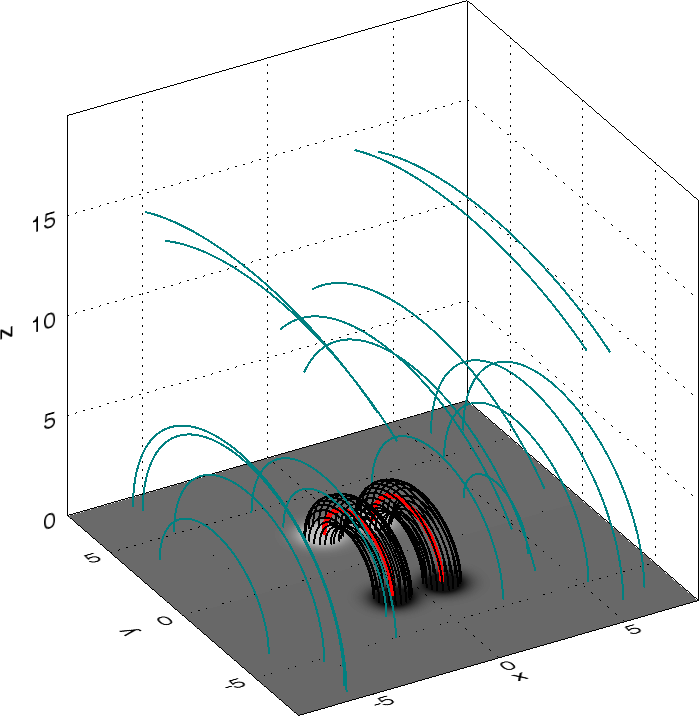}}}  
   \end{minipage}   
  \caption{Initial magnetic configuration and linkage of numerical simulation. \protect\subref{subfig:base000} illustrates the connectivity of field lines which begin in the right positive (R$^+$) or left positive (L$^+$) polarity regions, defined by $B_z\ge0.37$, with starting points coloured according to final positions of each field line, with those terminating at $x<0$ seen in blue and $x>0$ seen in red. Green contours indicate regions where $|B_z|=0.37$, while green crosses show the minimum value of $B_z$ in each negative region. \protect\subref{subfig:cfg000} shows interpolated magnetic field lines, illustrating the flux tubes (black), the axis of each (red) and ambient field (light blue).}
 \label{fig:t0}
\end{figure}

\subsection{Initial Configuration}
The initial (pre-shear) configuration can be seen in Figure~\ref{fig:t0}. The vertical magnetic field on the photospheric boundary consists of four sources, two positive and two negative, and each has the same form, $\pm B_0 \exp{(-r^2/r_0^2)}$, and flux. $r$ is a local radial coordinate centred on $x= \pm 1$ and $y =\pm 2.5$ and $r_0$ is a measure of the width of each source. The contours of vertical magnetic field strength in Figure~\ref{subfig:base000} illustrate the magnetic configuration imposed on the base of the simulation, while Figure~\ref{subfig:cfg000} shows the corresponding 3D magnetic field, illustrating that the configuration initially consists of two untwisted flux tubes. The field is left to relax until $t=1200$ by which time it is very close to potential. To simplify the discussion, we will label each of the strong polarity regions in Figure~\ref{subfig:base000} using the notation of Figure~\ref{fig:zip2}. L$^+$ signifies the region to the left of the line $x=0$ having positive polarity in $B_z$ (initially found centred on $[x,y]=[-1,2.5]$) while L$^-$ signifies the region of negative $B_z$ polarity to the left of $x=0$ (initially located at $[-1.-2.5]$). Similarly R$^+$ and R$^-$ signify the regions of positive or negative polarity in the positive $x$ domain (with R$^+$ initially centred on $[1,2.5]$ and R$^-$ initially centred on $[1,-2.5]$). Analysis of the initial connectivity linking these regions (seen in Figure~\ref{subfig:cfg000}) shows that field lines which originate in L$^+$(R$^+$) all terminate in L$^-$(R$^-$) respectively; there is no field linking the left and right halves of the domain in the initial configuration.  

\subsection{Shearing Motions}
From $t=1000$ to $t=1600$, we impose a slow, sub-Alfv\'enic shearing velocity on the photospheric base that switches on at $t=1200$ and off at $t=1400$. Thus, on $z=0$ and for $|x| < 2$, we impose
\begin{equation}
v_y(x,y,0, t) = 0.007\left \{\tanh\left (\frac{t-1200}{2}\right ) - \tanh\left (\frac{1400-t}{2}\right ) \right \}\sin \left (\frac{\pi x}{2}\right)\; .
\end{equation}
Both $v_x$ and $v_z$ remain zero on this boundary and all other variables have zero normal derivative. This shearing motion moves both sets of sources until L$^+$ is brought approximately opposite R$^-$. The source motion is not completely ideal; all four sources are slightly spread out during their movement. As L$^+$ and R$^-$ move closer together, the current density {above the photosphere} increases. From $t=1400$ until $t=1600$, the driving has stopped but the current {above the photosphere} now exceeds $j_{\rm{crit}}$ and the loops begin to reconnect. {Our experiment is primarily designed to study the reconnection of these super-critical ``coronal" currents. Our inclusion of a smaller background resistivity means that, in addition, we also anticipate a small amount of source diffusion and possibly flux cancellation when opposite sources are close together. While non-negligible, these effects will play a secondary role to the super-critical currents and the value of $\eta_{a}$ which drive zipper reconnection, and which dominate the behaviour revealed in this experiment.}

\section{Results}\label{sec:results}
We now examine various aspects of the results of this experiment.

\subsection{Evolution of Base Connectivity and $B_z$ Flux}\label{subsec:base}
We first assess how the connectivity of the sources evolves over time, both during and after the shearing phase of the experiment. In order to do so, we must first redefine the regions labelled L$^+$, L$^-$, R$^+$ and R$^-$. The initial positive and negative sources (seen in Figure~\ref{subfig:cfg000}) overlap. Distinct contours for sources in either half-plane can only be formed for values of $|B_z|>0.37$ (as shown in the original image); smaller values of $B_z$ are unable to distinguish between right and left sources of the same sign. For the time being, the positive flux source core regions are defined by the critical $B_z=0.37$ contours, while the negative source core regions are defined by $B_z=-0.37$. It should be noted that such definitions mean that our ``sources'' already omit some of the total flux imposed at the base.

\begin{figure}[t]
 \centering
  \subfloat[$t=1300\tau_{\rm{A}}$]{\label{subfig:t130}\resizebox{0.44\textwidth}{!}{\includegraphics[clip=true, trim=20 20 70 87]{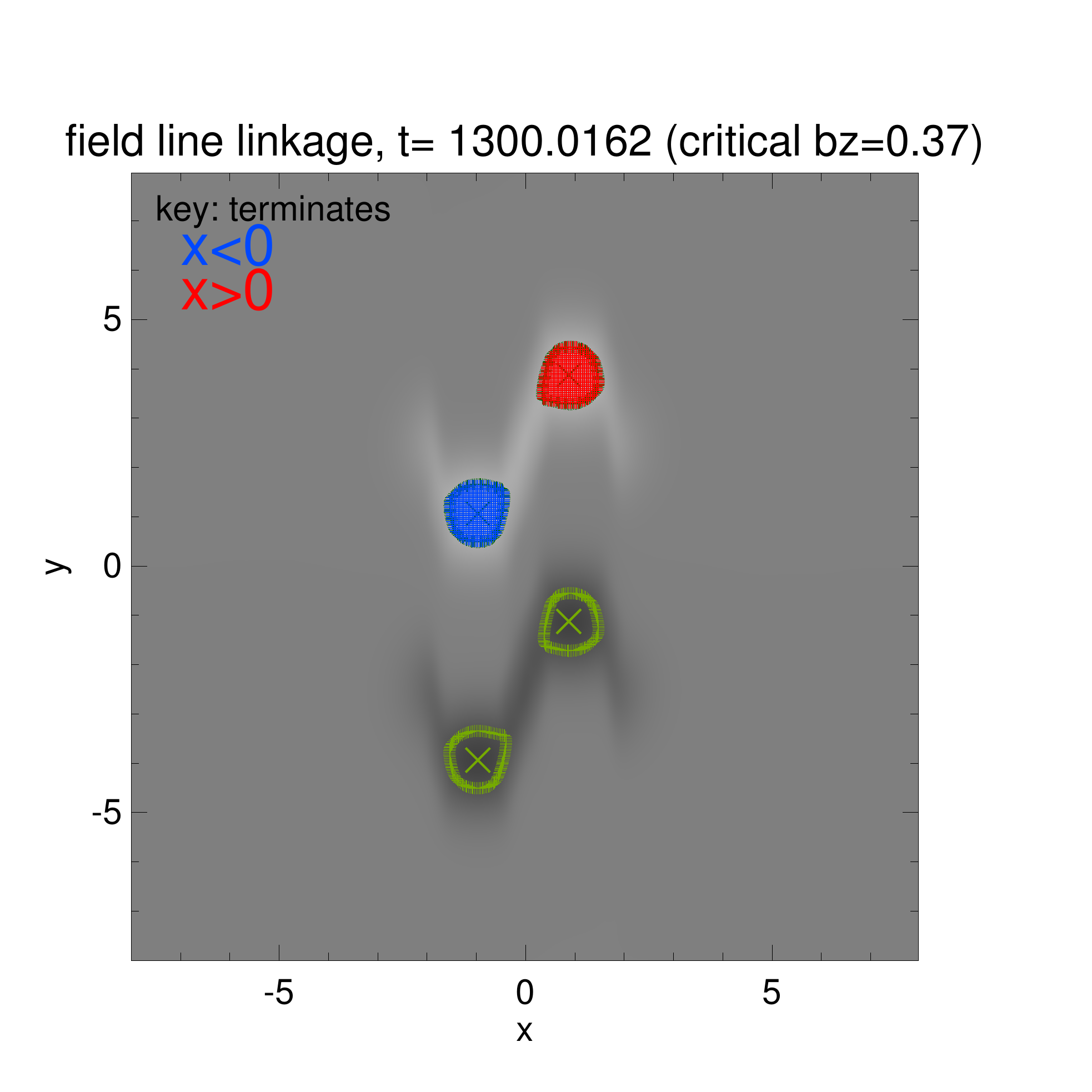}}}
  \subfloat[$t=1400\tau_{\rm{A}}$]{\label{subfig:t140}\resizebox{0.44\textwidth}{!}{\includegraphics[clip=true, trim=20 20 70 87]{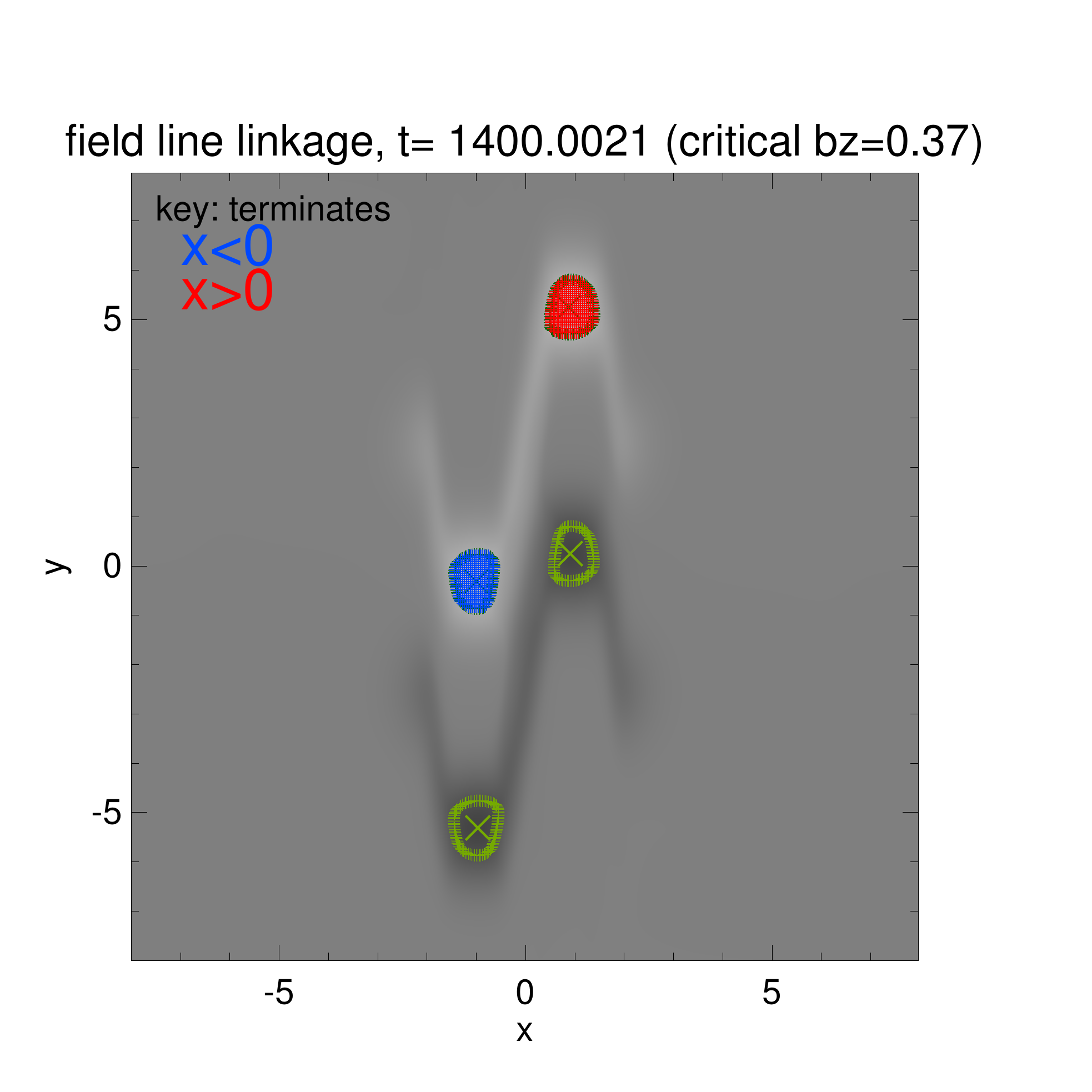}}}\hfill\resizebox{0.11\textwidth}{!}{\includegraphics{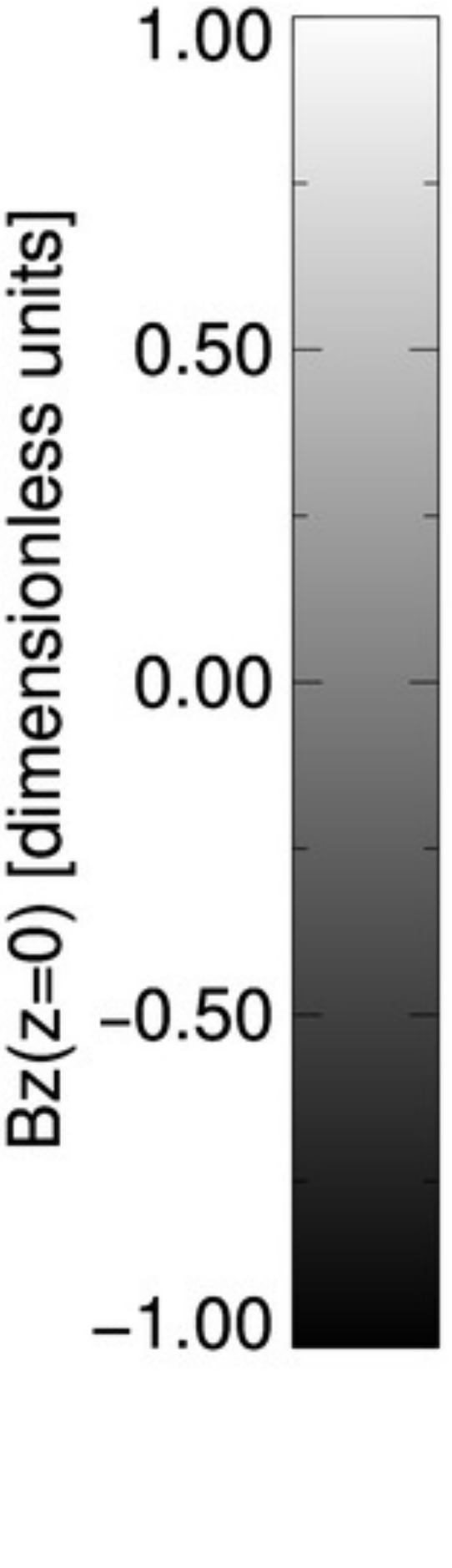}}\\
  \subfloat[$t=1500\tau_{\rm{A}}$]{\label{subfig:t150}\resizebox{0.44\textwidth}{!}{\includegraphics[clip=true, trim=20 20 70 87]{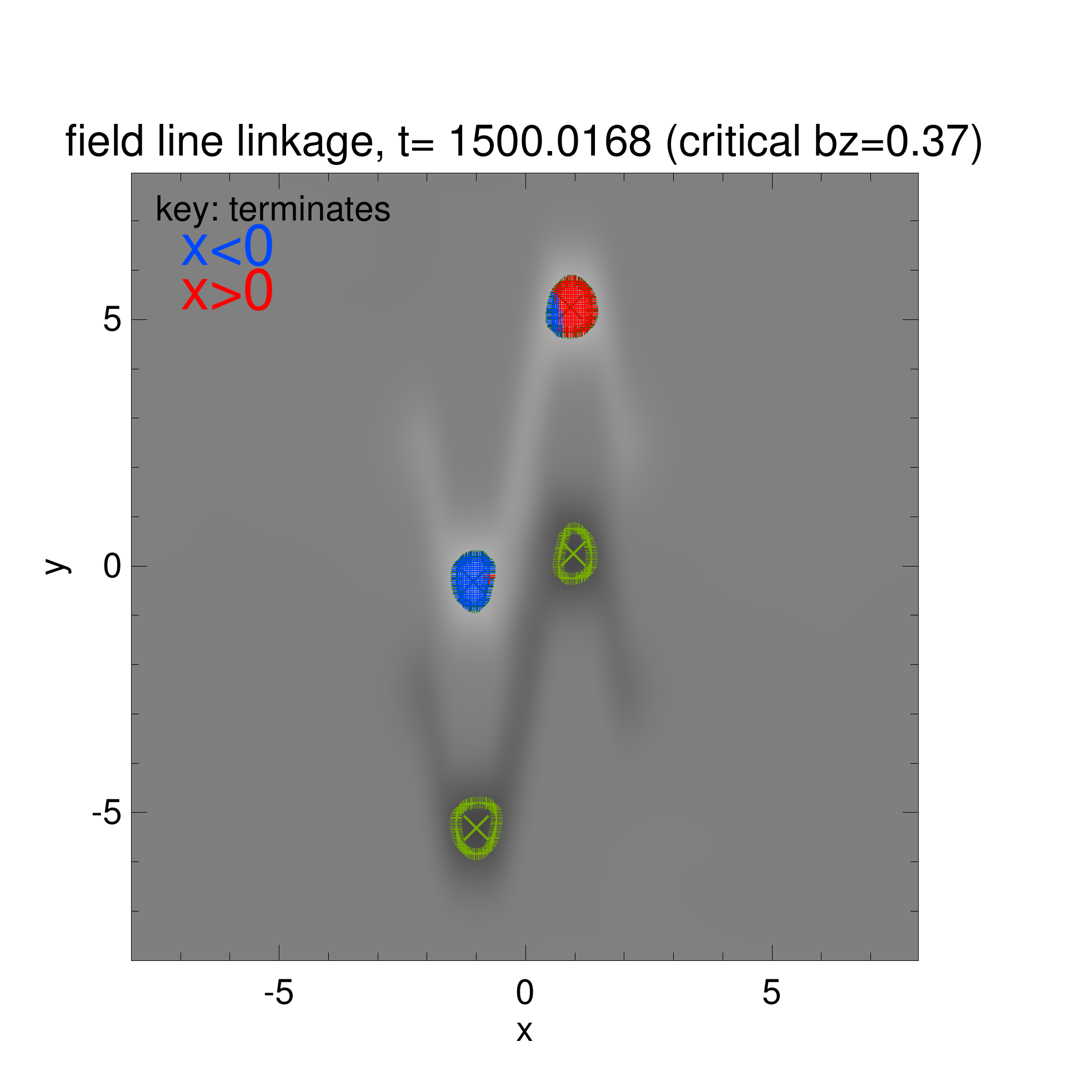}}}
  \subfloat[$t=1600\tau_{\rm{A}}$]{\label{subfig:t160}\resizebox{0.44\textwidth}{!}{\includegraphics[clip=true, trim=20 20 70 87]{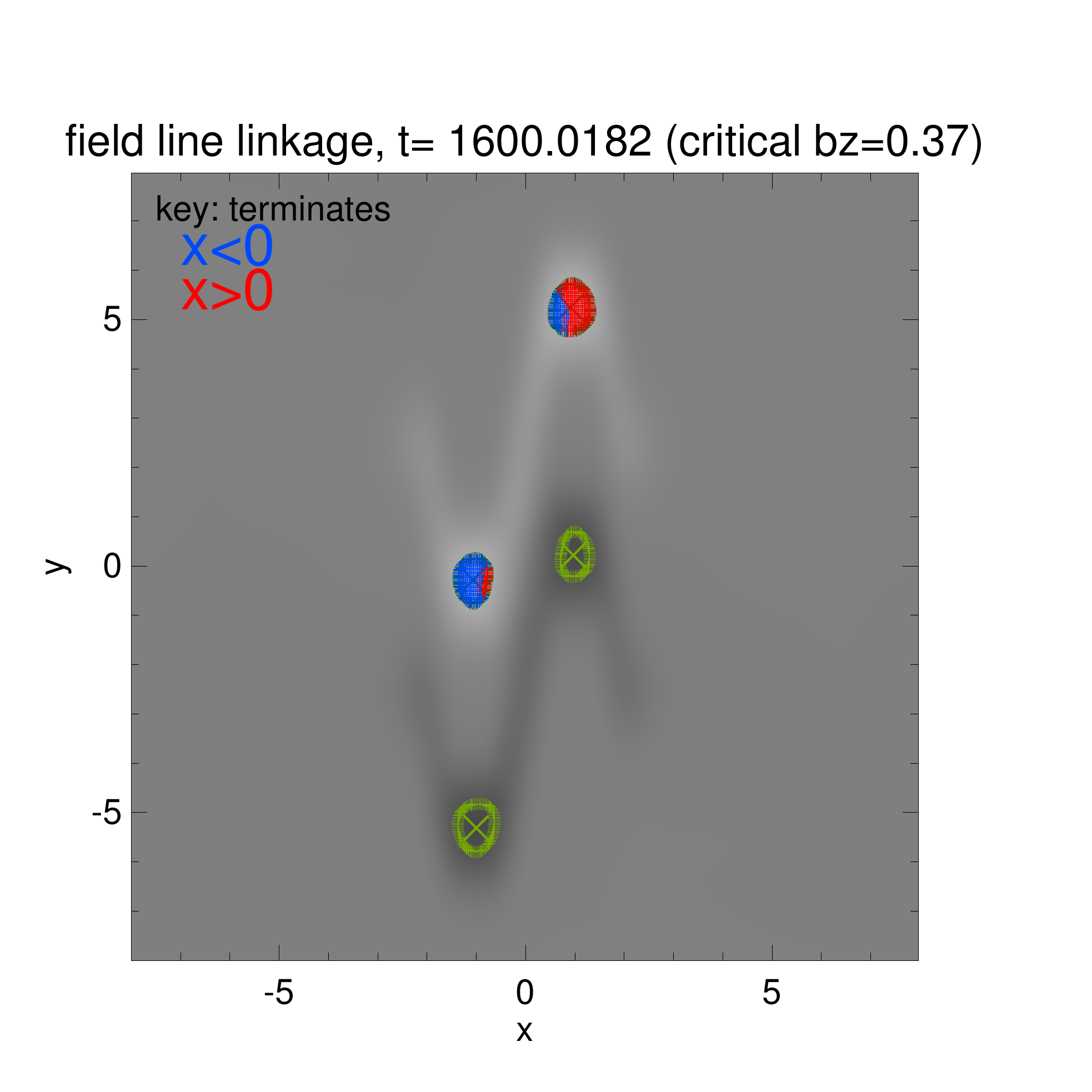}}}\hfill\resizebox{0.11\textwidth}{!}{\includegraphics{Bzcbar_vert-eps-converted-to.pdf}}
  \caption{Variation of connectivity over time (for critical $|B_z|=0.37$, shown by green contours). Positive contours of $B_z$ contain a grid of points for which field lines have been calculated, and coloured according to where each field line terminates, blue if $x<0$ and red if $x>0$. Green crosses indicate locations of minimum $B_z$ in each source.}
 \label{fig:conn37}
\end{figure}

To estimate how the connectivity of our sources evolves, we calculate a Cartesian grid of initial positions within each positive or negative sources, before tracing a magnetic field line through the domain from each point. If the final position of the field line lies in the left ($x<0$) half-plane, the initial position from which the field line is traced is coloured blue, while initial positions whose field lines terminate in the right ($x>0$) half-plane are coloured red. For brevity, we will only present the connectivity of field lines traced from the positive $B_z$ sources (L$^+$ and R$^+$) in Figure~\ref{fig:conn37}, at four different stages during and after the shearing phase of the experiment.

In Figure~\ref{fig:conn37}, we build up a picture of how the connectivity of each source changes with time. We can see that no connectivity changes are apparent during the shearing phase ($t=1300\tau_{\rm{A}}$, in Figure~\ref{subfig:t130}) or at the end of the shearing phase ($t=1400\tau_{\rm{A}}$, in Figure~\ref{subfig:t140}). However, by $t=1500\tau_{\rm{A}}$, Figure~\ref{subfig:t150} shows the existence of a large portion of the R$^+$ region coloured blue, meaning that links have formed with the left half-plane. A very small red inclusion is also notable in L$^+$; R$^+$ has much greater connectivity changes than L$^+$. A short time later,  ($t=1600\tau_{\rm{A}}$, in Figure~\ref{subfig:t160}), the blue inclusion in R$^+$ and the red inclusion in L$^+$ are much more similar in size. It is also apparent from Figure~\ref{fig:conn37} that the area surrounded by contours of $|B_z|=0.37$ is decreasing with time. Indeed, by $t=2000\tau_{\rm{A}}$, no flux is visible on the base that satisfies $|B_z|>0.37$; the polarity regions broaden over time, reducing the peak value of $B_z$ as they broaden.

To account for this effect, and to continue to monitor the connectivity of each of the source regions, we reduce the critical value of $B_z$ used to define the core of each source region. After $t=1600\tau_{\rm{A}}$, each strong polarity region is instead 
defined using contours of $|B_z|=0.2$. At this value, at $t=1600\tau_{\rm{A}}$, our sources once again remain distinct (weaker $B_z$ values would not reveal two positive and two negative sources) and broad (stronger $B_z$ values reduce the area of each source).  
\begin{figure}[t]
 \centering
  \subfloat[$t=1600\tau_{\rm{A}}$]{\label{subfig:t160b}\resizebox{0.44\textwidth}{!}{\includegraphics[clip=true, trim=20 20 70 87]{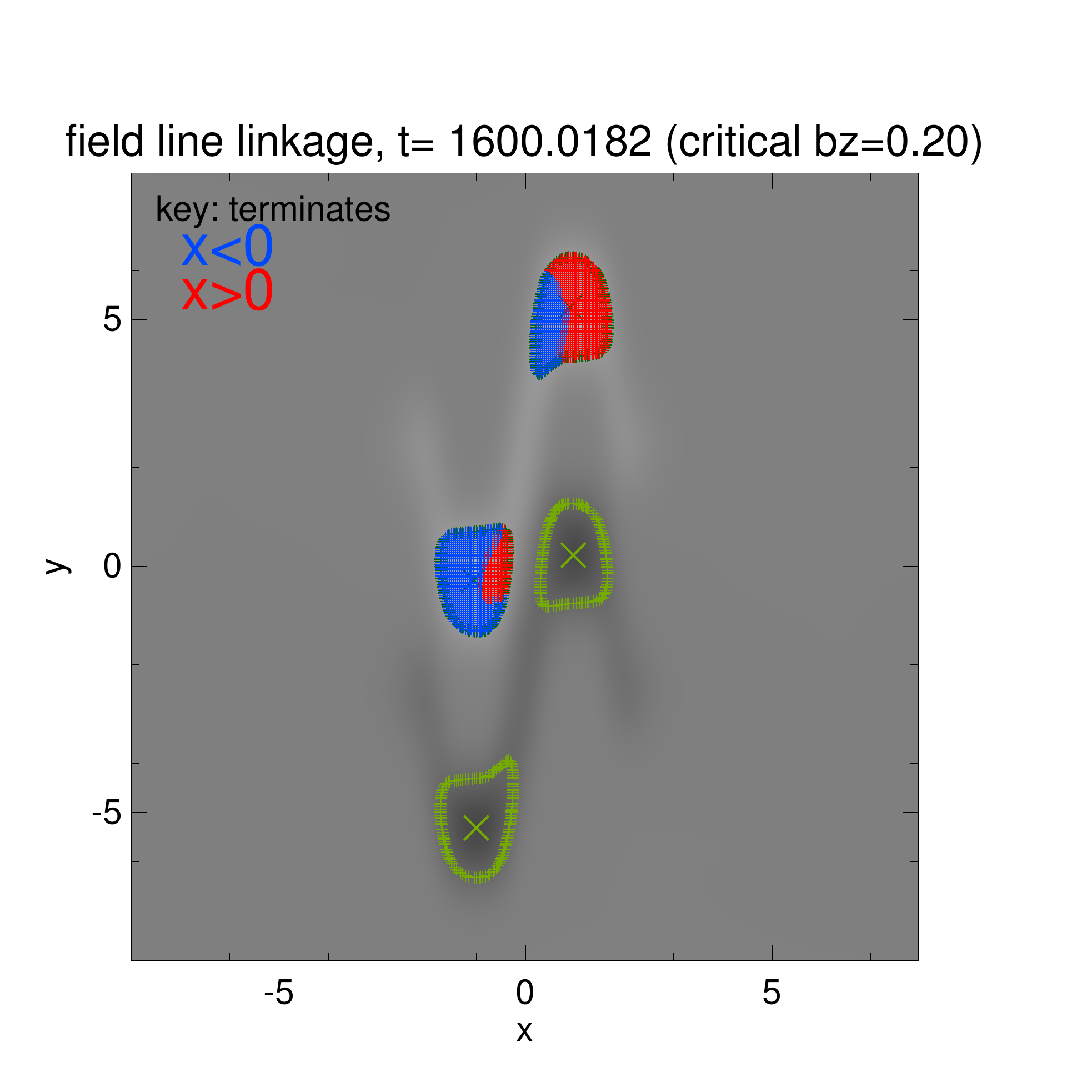}}}
  \subfloat[$t=2000\tau_{\rm{A}}$]{\label{subfig:t200}\resizebox{0.44\textwidth}{!}{\includegraphics[clip=true, trim=20 20 70 87]{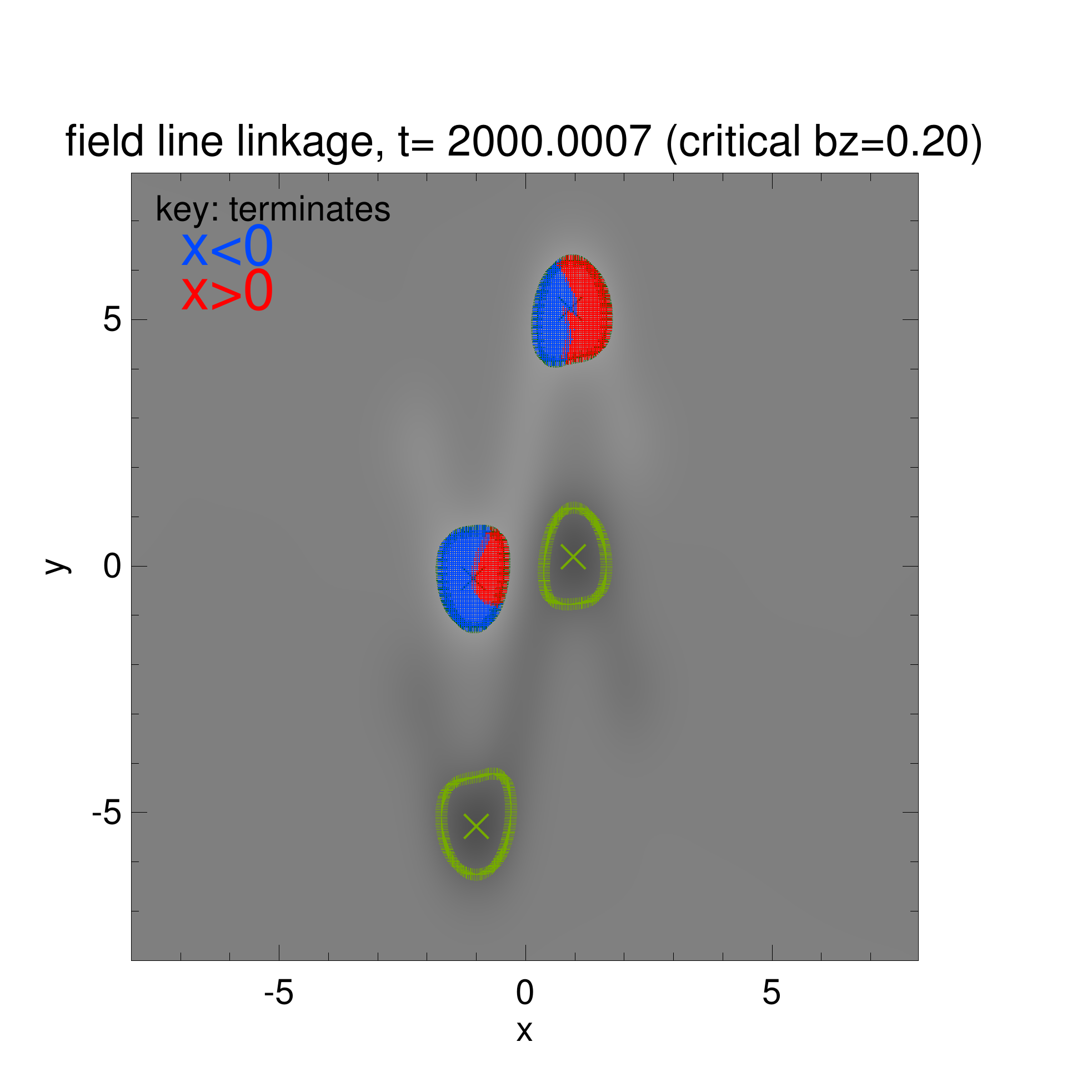}}}\hfill\resizebox{0.11\textwidth}{!}{\includegraphics{Bzcbar_vert-eps-converted-to.pdf}}\\
  \subfloat[$t=2400\tau_{\rm{A}}$]{\label{subfig:t240}\resizebox{0.44\textwidth}{!}{\includegraphics[clip=true, trim=20 20 70 87]{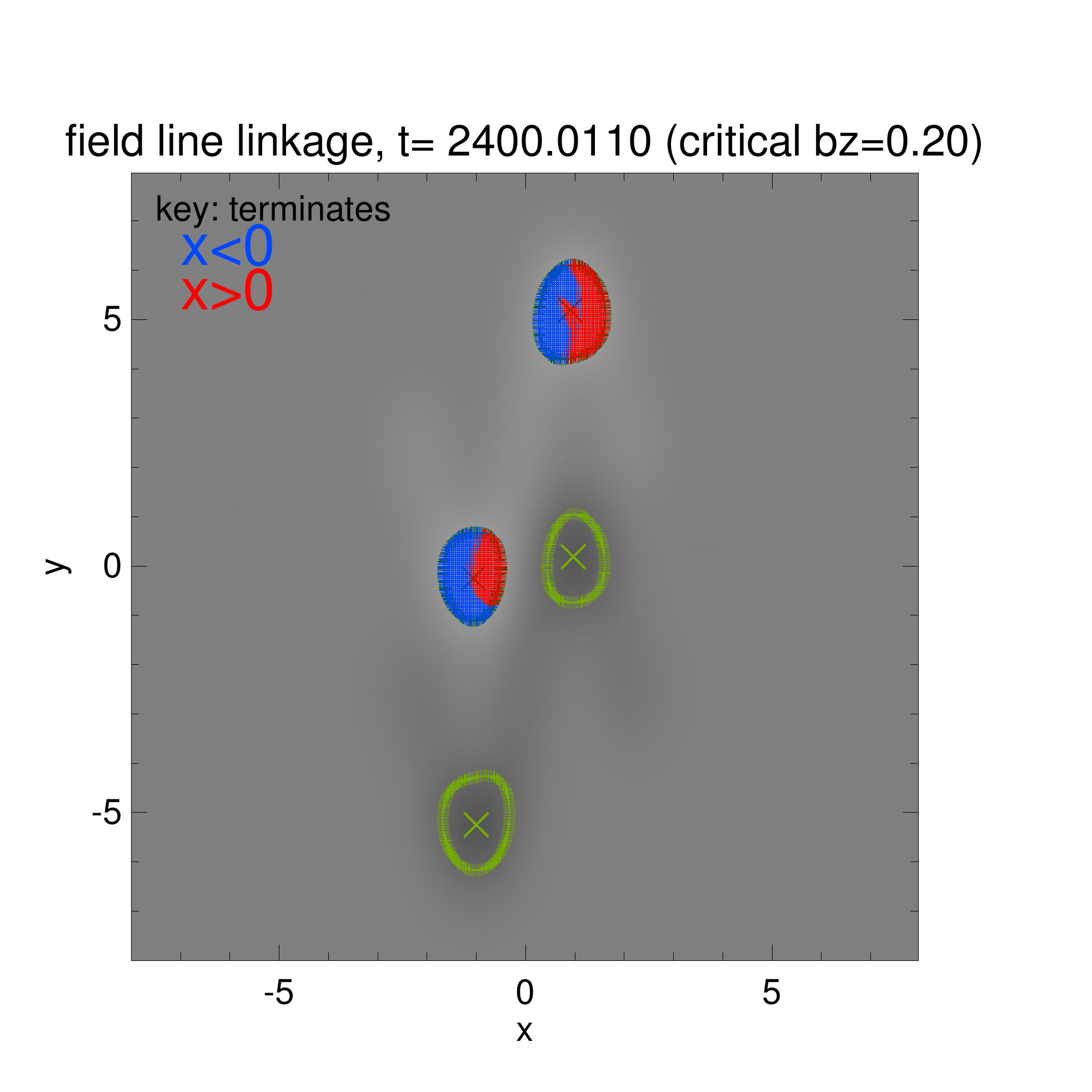}}}
    \subfloat[$t=2800\tau_{\rm{A}}$]{\label{subfig:t280}\resizebox{0.44\textwidth}{!}{\includegraphics[clip=true, trim=20 20 70 87]{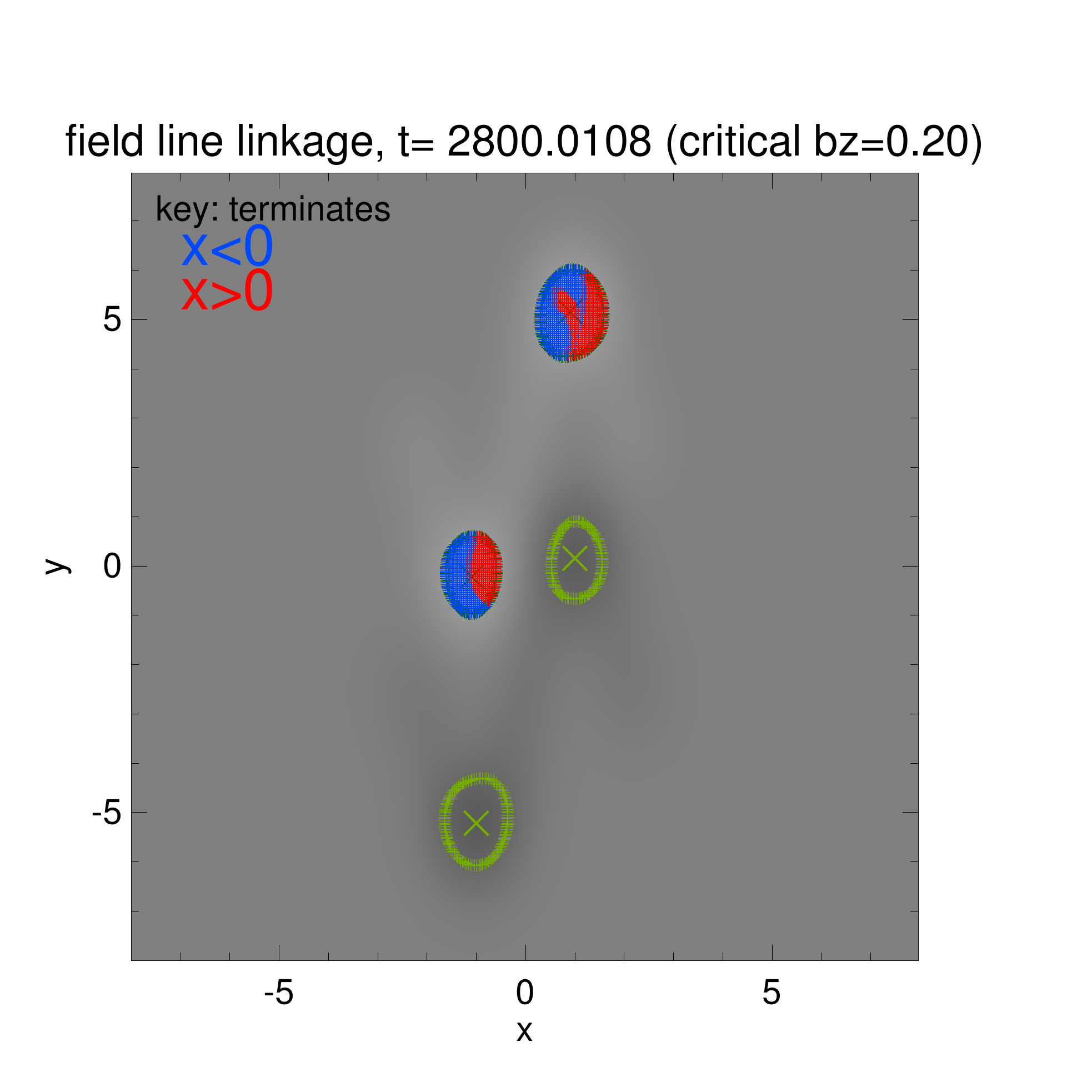}}}\hfill\resizebox{0.11\textwidth}{!}{\includegraphics{Bzcbar_vert-eps-converted-to.pdf}}
  \caption{Variation of connectivity over time (with a critical $B_z=0.2$, shown by green contours). Positive contours of $B_z$ contain a grid of points for which field lines have been calculated, and coloured according where each field line terminates, blue if $x<0$ and red if $x>0$. Green crosses indicate locations of minimum $B_z$ in each source.}
 \label{fig:conn20}
\end{figure}

The connectivity maps in Figure~\ref{fig:conn20} show that the sources at the base continue to diffuse at and after $t=1600\tau_{\rm{A}}$, and that the critical contours of $|B_z|=0.2$ omit large regions of weak positive and negative flux. Regarding the connectivity, we once again see that the blue incursion into R$^+$ is larger than the red incursion in L$^+$ at $t=1600\tau_{\rm{A}}$ and remains so throughout the remainder of the experiment. The connectivity map in R$^+$ also evolves somewhat differently to L$^+$: as the blue inclusion increases in area, a red feature appears to ``break off'' from the primary red region from $t=2000\tau_{\rm{A}}$. By the end of the experiment, most of the flux in the source which remains connected to the right half-plane is only found on the extreme right of R$^+$, while a separate ``island'' of red field lines has now formed near the centre of R$^+$ in Figure~\ref{subfig:t280}.

\begin{figure}[t]
  \subfloat[Critical $B_z=0.37$, $t=1300\rightarrow1600\tau_{\rm{A}}$]{\label{subfig:flux1}\resizebox{0.48\textwidth}{!}{\includegraphics[clip=true, trim=30 10 0 0]{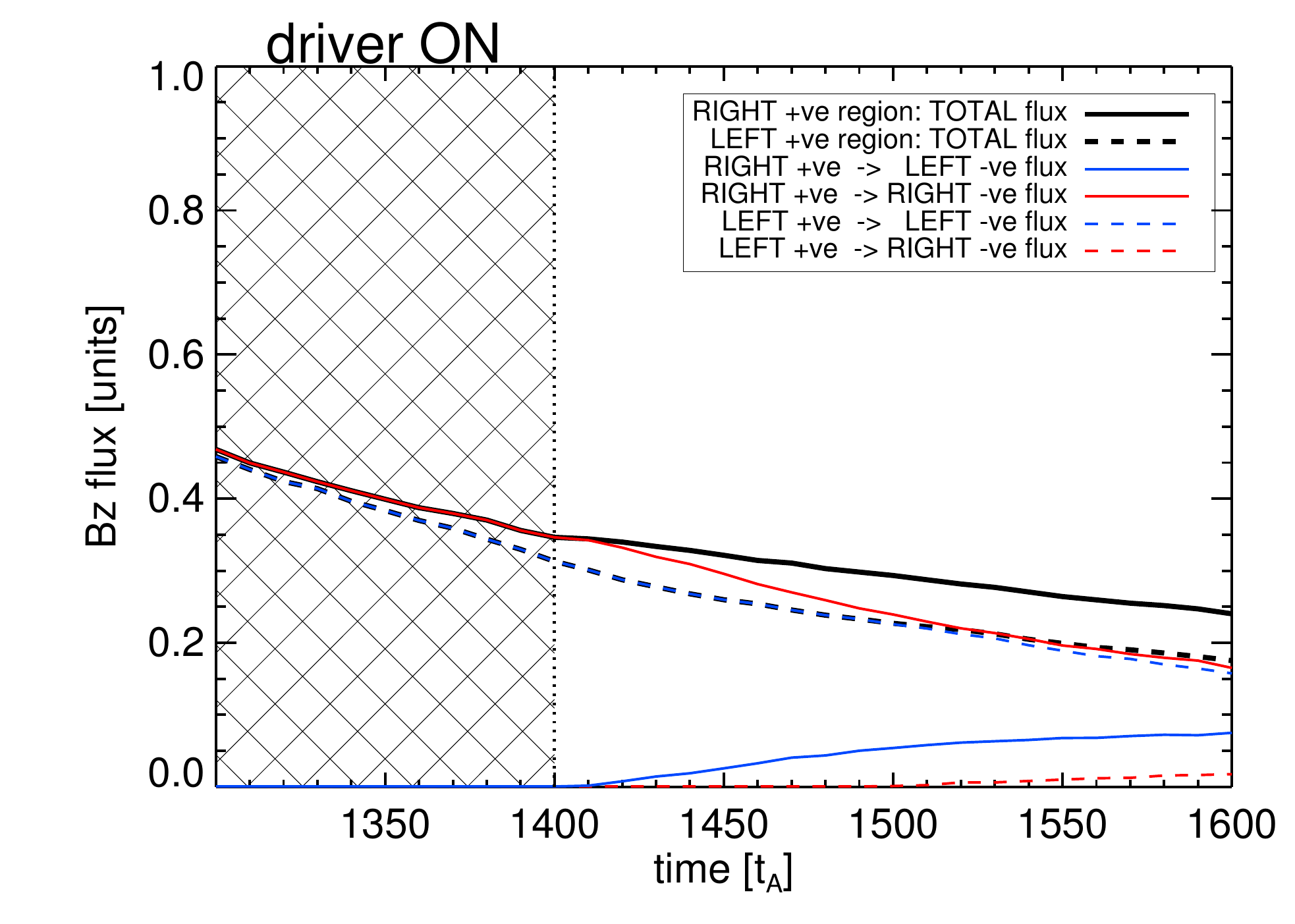} }}
  \subfloat[Critical $B_z=0.2$, $t=1600\rightarrow3000\tau_{\rm{A}}$]{\label{subfig:flux2}\resizebox{0.48\textwidth}{!}{\includegraphics[clip=true, trim=30 10 0 0]{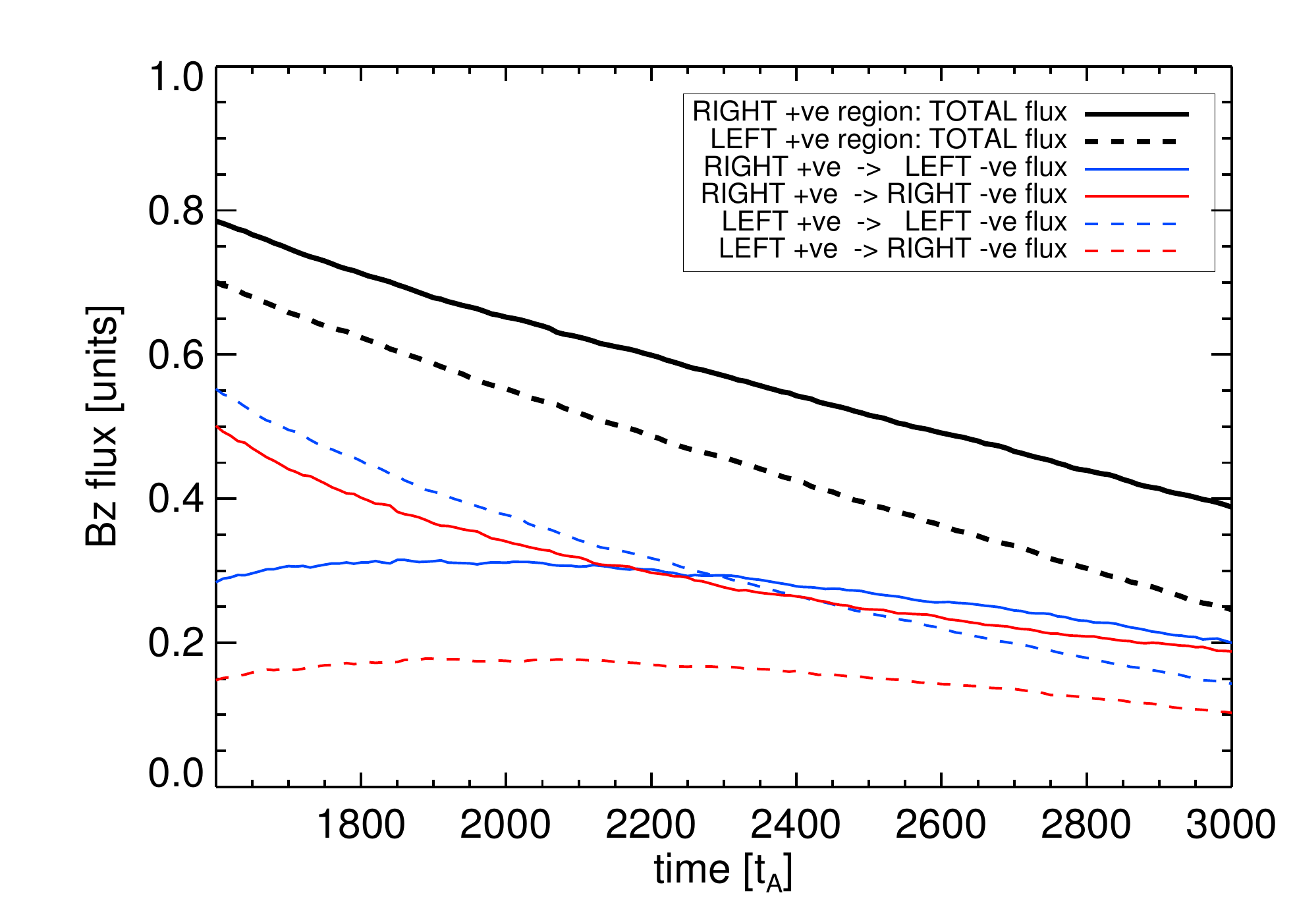}} }
  \caption{Variation of the magnetic flux through the positive polarity regions. Solid lines illustrate flux at the right positive source, dashed lines show flux from the left positive source. Colours show total (black), left (blue) and right (red) connected flux components (NB black line always equals sum of red and blue lines). The hatched region indicates times when velocity driver at base was activated. Fluxes are given in dimensionless units and can be related to real units through normalisation; for the quantities described earlier, one dimensionless unit of magnetic flux is equivalent to $1\times10^{11}$Wb or $1\times10^{19}$Mx.}
 \label{fig:flux20}
\end{figure}

Another quantity of interest is the amount of magnetic flux through each source, and how much flux links which sources. In order to estimate this, we evaluate the vertical magnetic field $B_z$ at each location on the grid (previously used to perform field line tracing). To estimate the total flux through each source, we sum $B_z$ in $x$ and $y$ directions. We also estimate the total flux which ultimately links to either right or left half-planes, only summing values in areas with field lines which link to $x>0$ or to $x<0$. We display the total flux (and its components, linked either to the right or left half-planes) in Figure~\ref{fig:flux20}, comparing the flux through $|B_z|=0.37$ contours in Figure~\ref{subfig:flux1} and the flux through $|B_z|=0.2$ contours in Figure~\ref{subfig:flux2}.

There are several noteworthy aspects of Figure~\ref{fig:flux20}. As expected, the total flux through both R$^+$ and L$^+$ decreases with time. The total flux through R$^+$ is greater than the total flux through L$^+$ during the driving phase. Also during this phase (and also illustrated by Figures~\ref{subfig:t130}--\ref{subfig:t140}), the total flux in L$^+$ is entirely linked to the left half-plane, while the total flux in R$^+$ is entirely linked to the right half-plane. This changes almost immediately after the driving phase ends: flux from R$^+$ terminating at $x<0$ rises from zero at approximately $1410\tau_{\rm{A}}$, matched by a corresponding drop in flux linking R$^+$ with $x>0$. Evidence of changing connectivity in L$^+$ occurs much later, with Figure~\ref{subfig:flux1} demonstrating an increase from zero of L$^+$ flux linked to $x>0$ at approximately $1510\tau_{\rm{A}}$, with a corresponding decrease in L$^+$ flux linkage with $x<0$.

Turning to the evolution shown in Figure~\ref{subfig:flux2}, decreasing the critical value of $|B_z|$ used to define each source region widens the area covered by each source and therefore increases the flux values recovered (relative to Figure~\ref{subfig:flux1}, which uses $|B_z|=0.37$ to define each source). It can be seen that the dominant connectivity of the R$^+$ region changes at later times: prior to $t\approx2100\tau_{\rm{A}}$, most of the total flux through R$^+$ terminates in the right half-plane. From $t\approx2100\tau_{\rm{A}}$ until the end of the experiment, the majority of flux through R$^+$ now links to the left half-plane. Such an effect also occurs in the flux through L$^+$, whose connectivity with the left half-plane always remains greater than the connectivity with the right half-plane. It is also worth noting that the connectivity of both sources appears to be tending towards equipartition, approaching equal levels of flux through each source linking to the left and right half-planes.

\begin{figure}[t]
\centering
 \resizebox{0.99\textwidth}{!}{\includegraphics{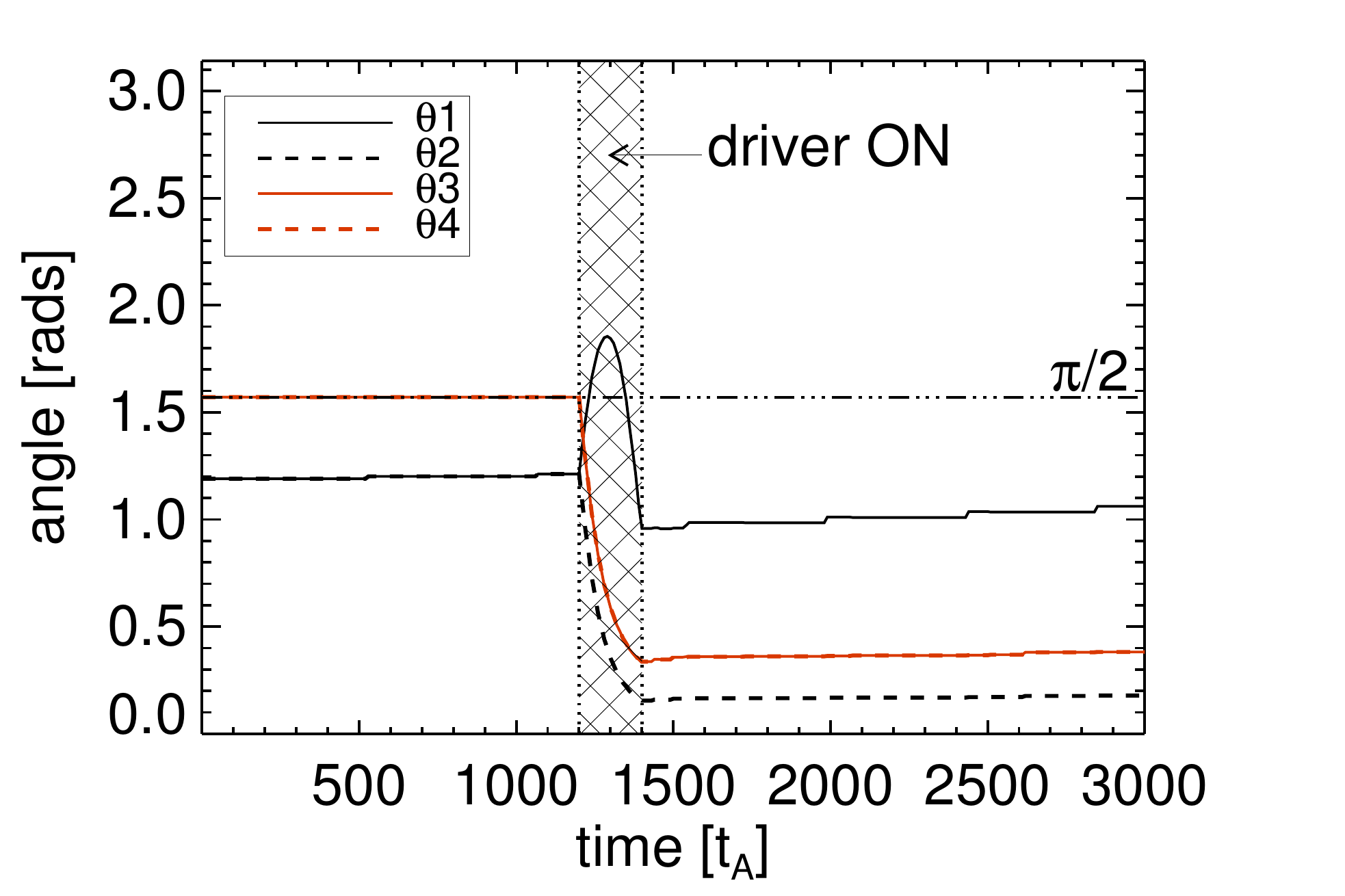}}
  \caption{The time evolution of the angles between positive and negative sources of vertical magnetic flux at the base of the simulation domain, for each angle defined in Figure~\ref{fig:zip2}. The hatched region indicates the period when velocity driver at base was activated, generating the shear responsible for reconnection.}
 \label{fig:theta}
\end{figure}

Finally, we present a comparison of the angles between each of the specific source regions, as defined in Figure~\ref{fig:zip2}, in order to estimate the magnetic helicity. Each angle is defined between the locations of peak positive and negative $B_z$ flux on the base of the simulation domain. These locations are not affected by any choice of critical $B_z$ threshold used to outline the sources. The evolution of the angles between these locations is seen in Figure~\ref{fig:theta}.

The figure shows that $\theta_1$ and $\theta_2$ are identical until the velocity driver is switched on, while $\theta_3$ and $\theta_4$ remain identical for all time. When the velocity driver is switched on, $\theta_1$ increases rapidly, rising above $\pi/2$ before reducing to a narrower angle than seen before the driving phase. The rise and fall can be attributed to the L$^+$ source region moving relative to R$^+$ and R$^-$. The angle $\theta_1$ is widest when L$^+$ lies at a $y$-value halfway between R$^+$ and R$^-$ (as approximately shown in Figure~\ref{subfig:t130}). L$^+$ continues to move down below $y=0$, while R$^+$ and R$^-$ move upwards in $y$, causing $\theta_1$ to become much narrower by the end of the driving phase (seen \textit{e.g.} in Figure~\ref{subfig:t160}). Meanwhile, according to Figure~\ref{fig:theta}, $\theta_2$ (and indeed $\theta_3$ and $\theta_4$) reduces over the course of the driving phase, as the angle subtended by R$^+$, L$^-$ and R$^-$ closes and the sources are progressively sheared.

\subsection{Overlying Flux Rope: Formation}\label{subsec:form}
\begin{figure}[t]
 \centering
 \subfloat[$t=1600\tau_{\rm{A}}$]{\label{subfig:cfg160}\resizebox{0.49\textwidth}{!}{\includegraphics[clip=true, trim=15 30 65 40]{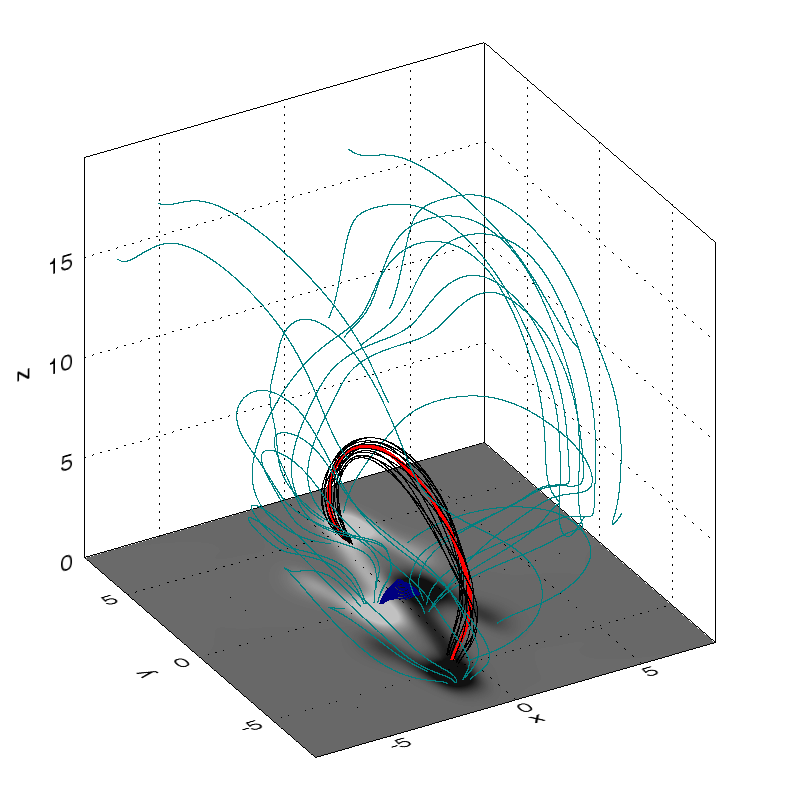}}}
 \subfloat[$t=3000\tau_{\rm{A}}$]{\label{subfig:cfg300}\resizebox{0.49\textwidth}{!}{\includegraphics[clip=true, trim=35 30 45 40]{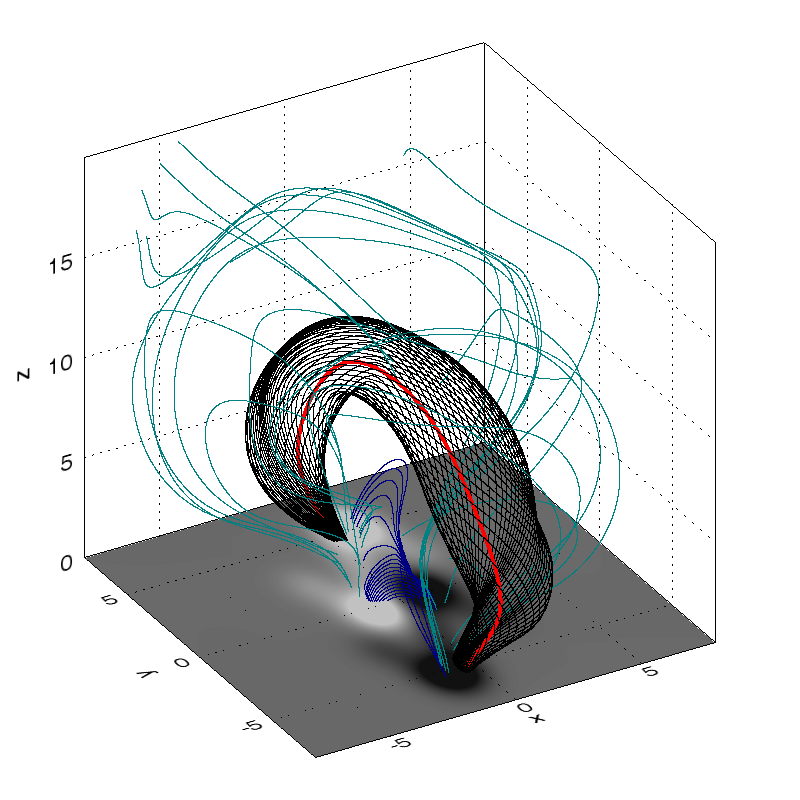}}}
  \caption{3D magnetic field configuration \protect\subref{subfig:cfg160} shortly after the end of the shearing phase of the experiment and \protect\subref{subfig:cfg300} at a much later stage of development of the flux rope. Images show magnetic field lines linking a pair of regions of opposite magnetic polarity of $B_z$ at the base, with black field lines illustrating the presence of a flux tube (with an axis indicated in {red}), {blue} field lines indicating {reconnected} field beneath the flux tube and {light blue} lines indicating ambient field, }
 \label{fig:ropeformation}
\end{figure}

One of the goals of this experiment was to establish if this process could lead to the formation of new flux ropes in the manner described in the zipper model \citep{paper:PriestLongcope2017}. By tracing selected magnetic field lines in the domain, it is easy to see that a new twisted flux rope does indeed form between sources R$^+$ and L$^-$ after the shearing phase of the experiment ends. As shown in Figure~\ref{fig:ropeformation}, this flux rope expands to fill the simulation domain over time, and overlies an arcade of magnetic field lines which link L$^+$ and R$^-$.

\subsection{Overlying Flux Rope: Twist and Helicity}\label{subsec:twist}
\begin{figure}[t]
 \centering 
  {\resizebox{0.7\textwidth}{!}{\includegraphics[clip=true, trim=0 0 0 0]{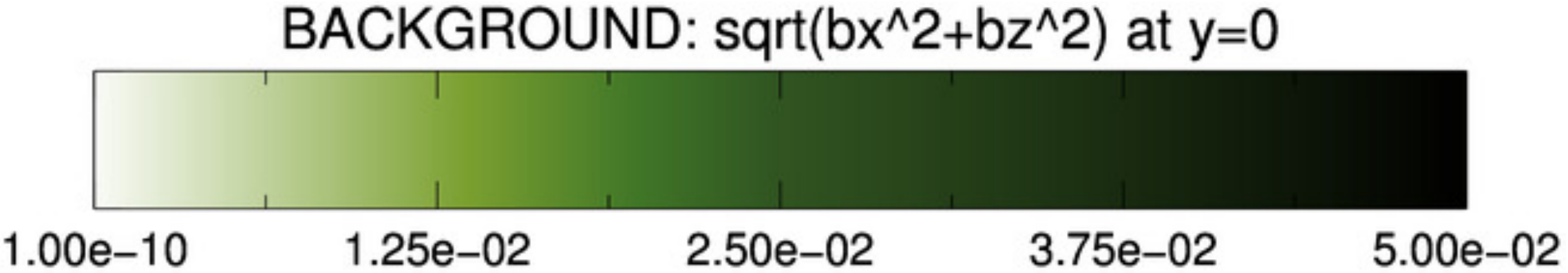}}}
  \subfloat[$t=1600\tau_{\rm{A}}$]{\label{subfig:2dtwist160}\resizebox{0.31\textwidth}{!}{\includegraphics[clip=true, trim=0 30 0 60]{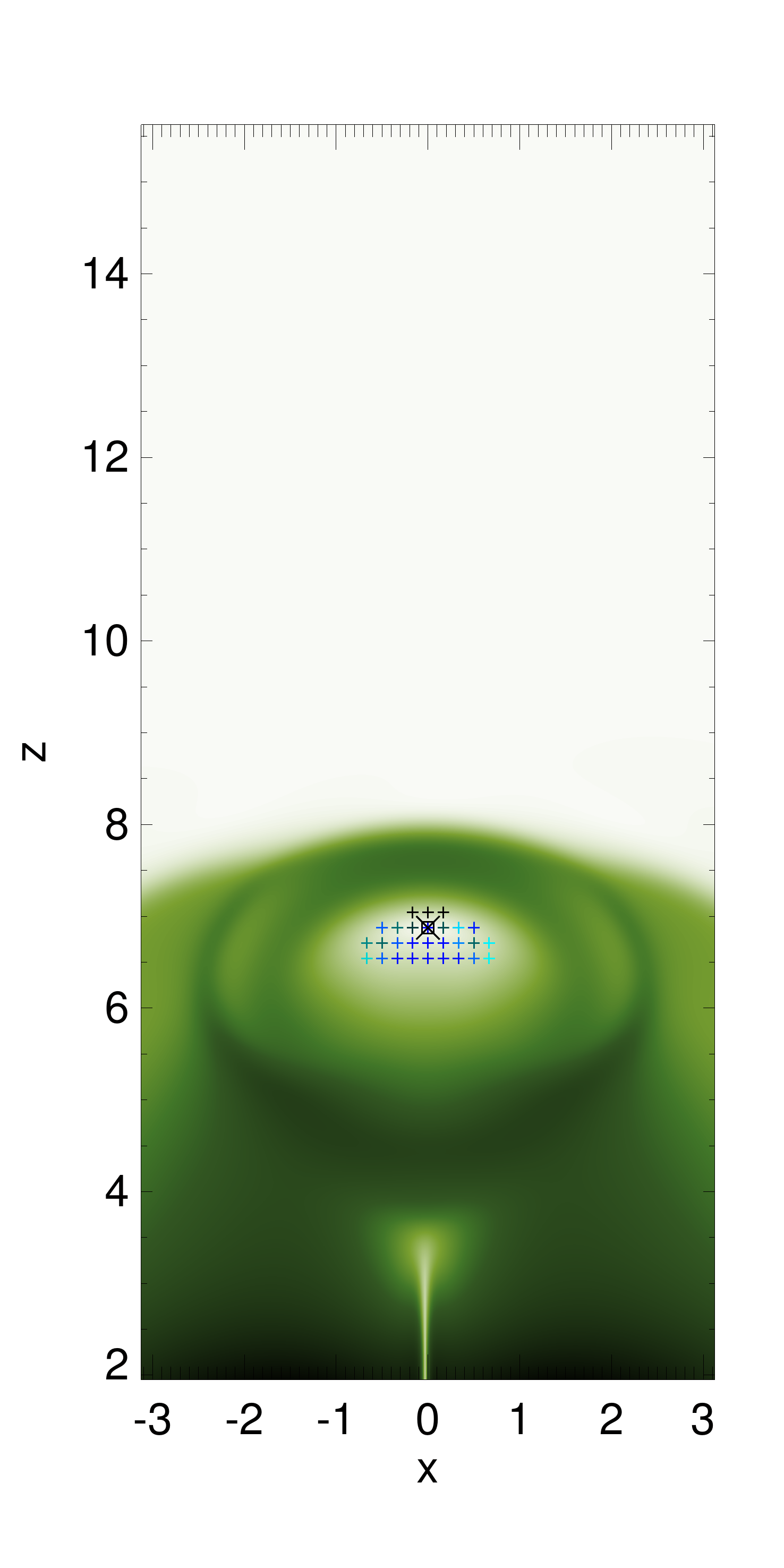}}}
  \subfloat[$t=1800\tau_{\rm{A}}$]{\label{subfig:2dtwist180}\resizebox{0.28\textwidth}{!}{\includegraphics[clip=true, trim=40 30 0 60]{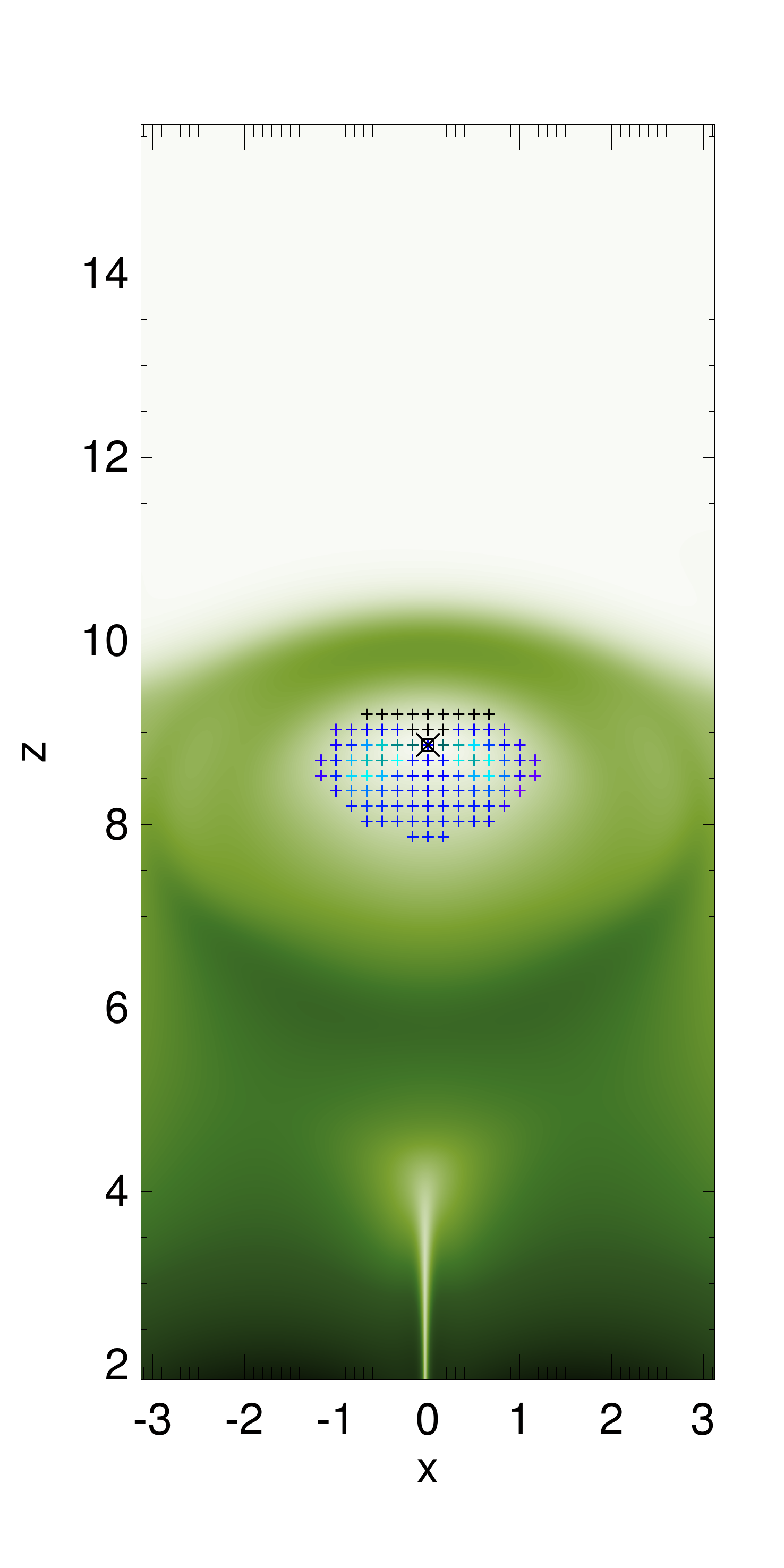}}}
  \subfloat[$t=3000\tau_{\rm{A}}$]{\label{subfig:2dtwist300}\resizebox{0.28\textwidth}{!}{\includegraphics[clip=true, trim=40 30 0 60]{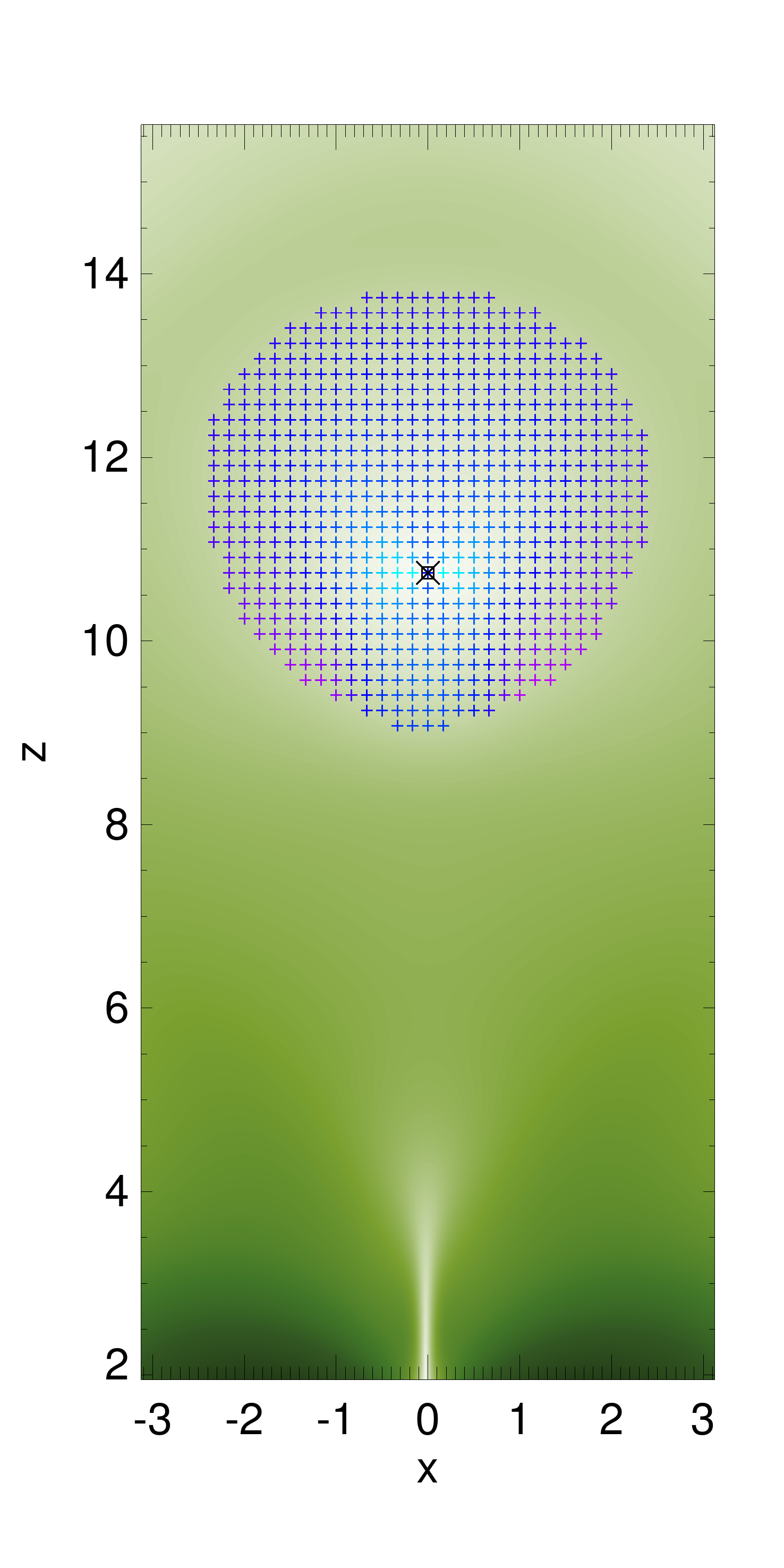}}}
   {\resizebox{0.12\textwidth}{!}{\includegraphics[clip=true, trim=0 -40 0 0]{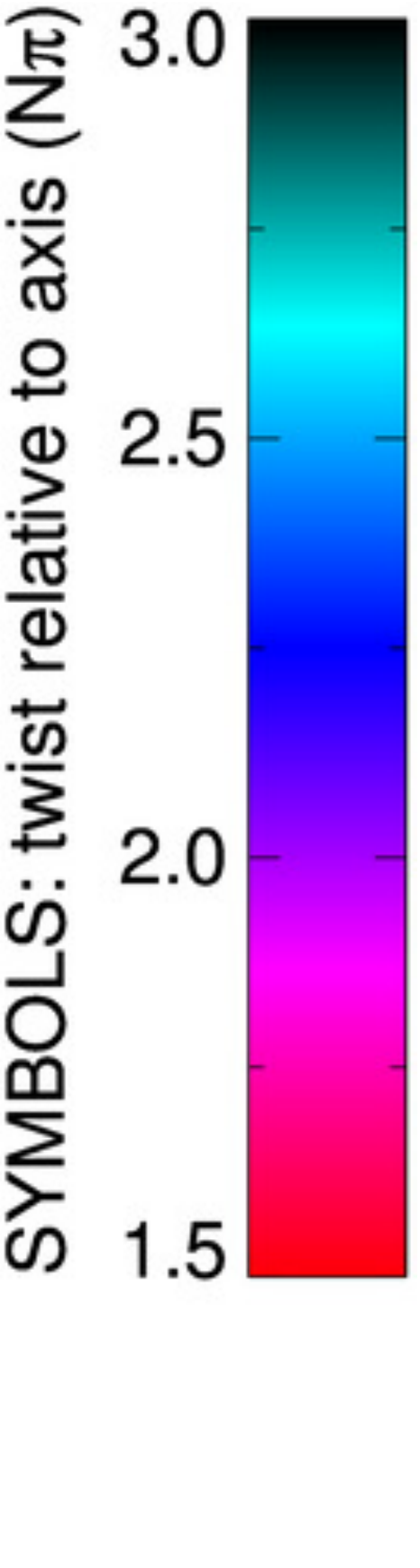}}}
  \caption{The distribution of twist in a cross-section of an evolving flux rope. The background of each image illustrates the perpendicular component of magnetic field, used to define the flux rope \protect\subref{subfig:2dtwist160} shortly after the initial shear ($1600\tau_{\rm{A}}$), \protect\subref{subfig:2dtwist180} as the flux rope expansion begins ($1800\tau_{\rm{A}}$) and \protect\subref{subfig:2dtwist300} at a much more developed stage ($3000\tau_{\rm{A}}$). The symbols indicate locations where the magnetic field has been interpolated, colour coded with the amount of twist relative to the axial field line (black) between the loop footpoints.}
 \label{fig:2dtwistcompare}
\end{figure}

The field lines which form the overlying flux rope in Figure~\ref{fig:ropeformation} are twisted, forming a helical path around a central axis. A key question is what is the twist of these field lines (relative to the flux rope axis) and how is the twist distributed throughout the tube? To calculate the twist at any location in the tube, we again create a grid of initial positions, this time distributed in the $x z$-plane at $y=0$. The grid is restricted to lie within a chosen contour of the perpendicular magnetic field strength (\textit{i.e.} $\sqrt{B_x^2+B_z^2}$) which we use to define our flux rope cross-section. The grid is used to trace field lines forwards (to one negative flux source at the base) and backwards (to the positive source), with the recovered locations at the base then used to calculate an angle of rotation relative to the flux rope axis (defined where $\sqrt{B_x^2+B_z^2}=0$ in the $[x,y=0,z]$ plane). In Figure~\ref{fig:2dtwistcompare}, we have colour-coded each position on each grid to represent the angle through which that field line rotates around the flux rope axis in three different snapshots as the flux rope rises and expands (and have included the perpendicular field strength in the background for context).

From Figure~\ref{fig:2dtwistcompare}, we see that the majority of field lines within the flux rope are twisted by between $1.9-2.5\pi$. The highest values of twist form in two thin bands, which broaden over time. By the end of the experiment,  
the peak values of twist are focussed close to the axis of the flux rope, but small regions of weaker twist ($<1.7\pi$) have become apparent at the edges of the flux tube.

The twist of field lines which form the flux rope (about its axis) is a crucial component of magnetic helicity. The conservation of total magnetic helicity is a natural consequence of magnetic flux tube formation and eruptions, since
it is conserved during three-dimensional reconnection \citep[see \textit{e.g.}][]{paper:Priestetal2016}. The total magnetic helicity ($H_{\rm{T}}$) can be expressed as the sum of the self-helicity ($H_{\rm{s}}$) resulting from the twist of each flux tube in the system and the mutual helicity ($H_{\rm{M}}$) of each tube relative to each other; for a system containing $N$ tubes, this amounts to:
\begin{equation}
H_{\rm{T}}=\sum_{i=1}^N H_{\rm{s}}^i+2\sum_{i<j=1}^N H_{\rm{M}}^{ij}.
\end{equation}
Hence, in a system containing two tubes labelled ${\rm{A}}$ and ${\rm{B}}$, this becomes
\begin{equation}
H_{\rm{T}}=H_{\rm{s}}^{\rm{A}}+H_{\rm{s}}^{\rm{B}}+H_{\rm{M}}^{\rm{AB}}.
\end{equation}
The self-helicity of any flux rope can be calculated as a product of the twist ($\phi$) and the magnetic flux $F$ passing through the tube (with a perpendicular cross-section $S$). For simplicity, we will use the definition outlined in \citet{paper:Priestetal2016}, using the mean twist of the flux tube $\bar{\phi}$, namely,
\begin{equation}
H_{\rm{s}}=\frac{\bar{\phi}}{2\pi}F^2, \qquad\qquad F=\oint_{\rm{s}} {\bf{B}}\cdot{\bf{ds}}.
\end{equation}

\begin{figure}[t]
 \centering
  \subfloat[Average twist]{\label{subfig:Avetwist}\resizebox{0.46\textwidth}{!}{\includegraphics[clip=true, trim=60 0 20 20]{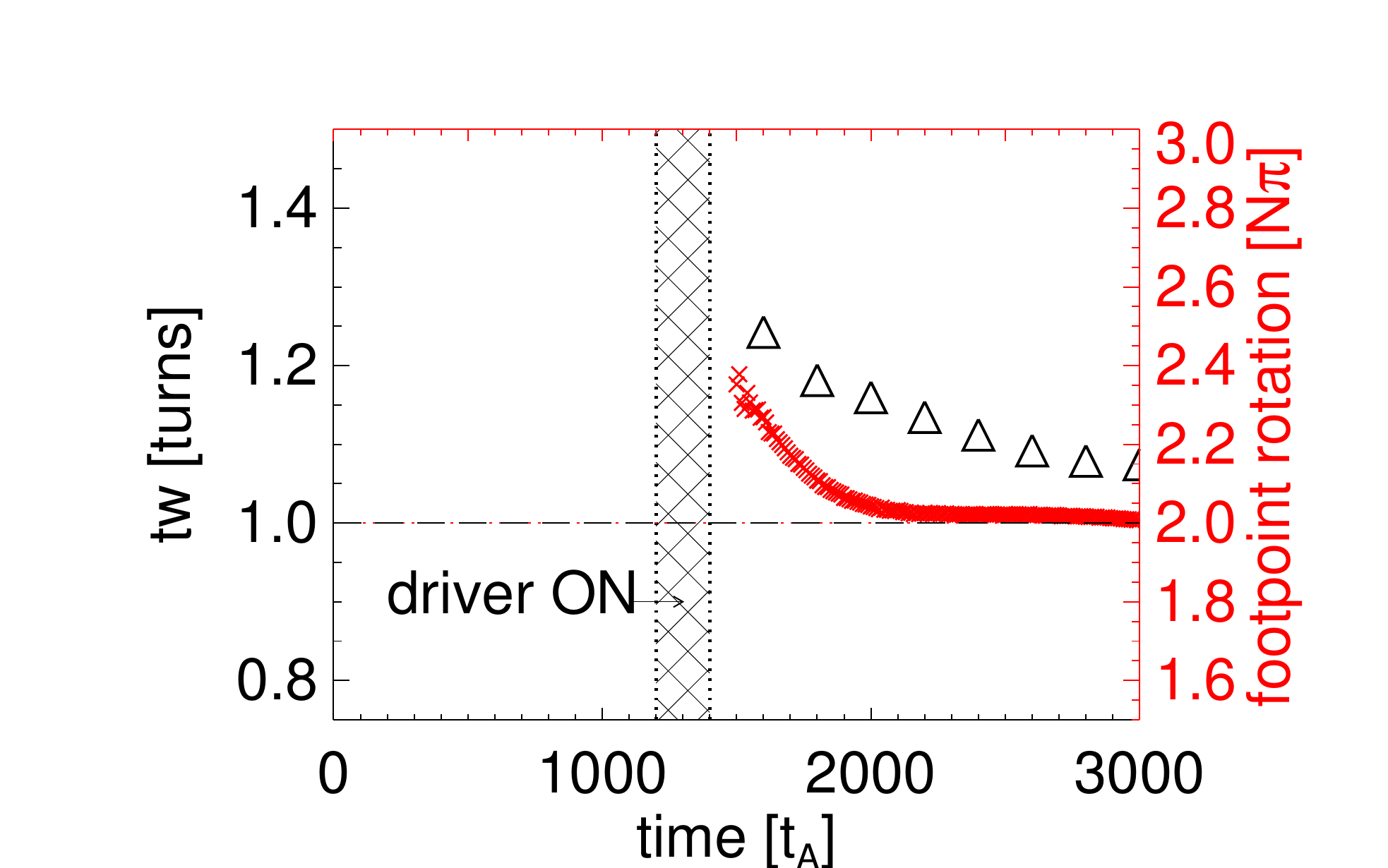}}}
  \subfloat[Self helicity]{\label{subfig:SelfHelicity}\resizebox{0.53\textwidth}{!}{\includegraphics[clip=true, trim=0 0 20 20]{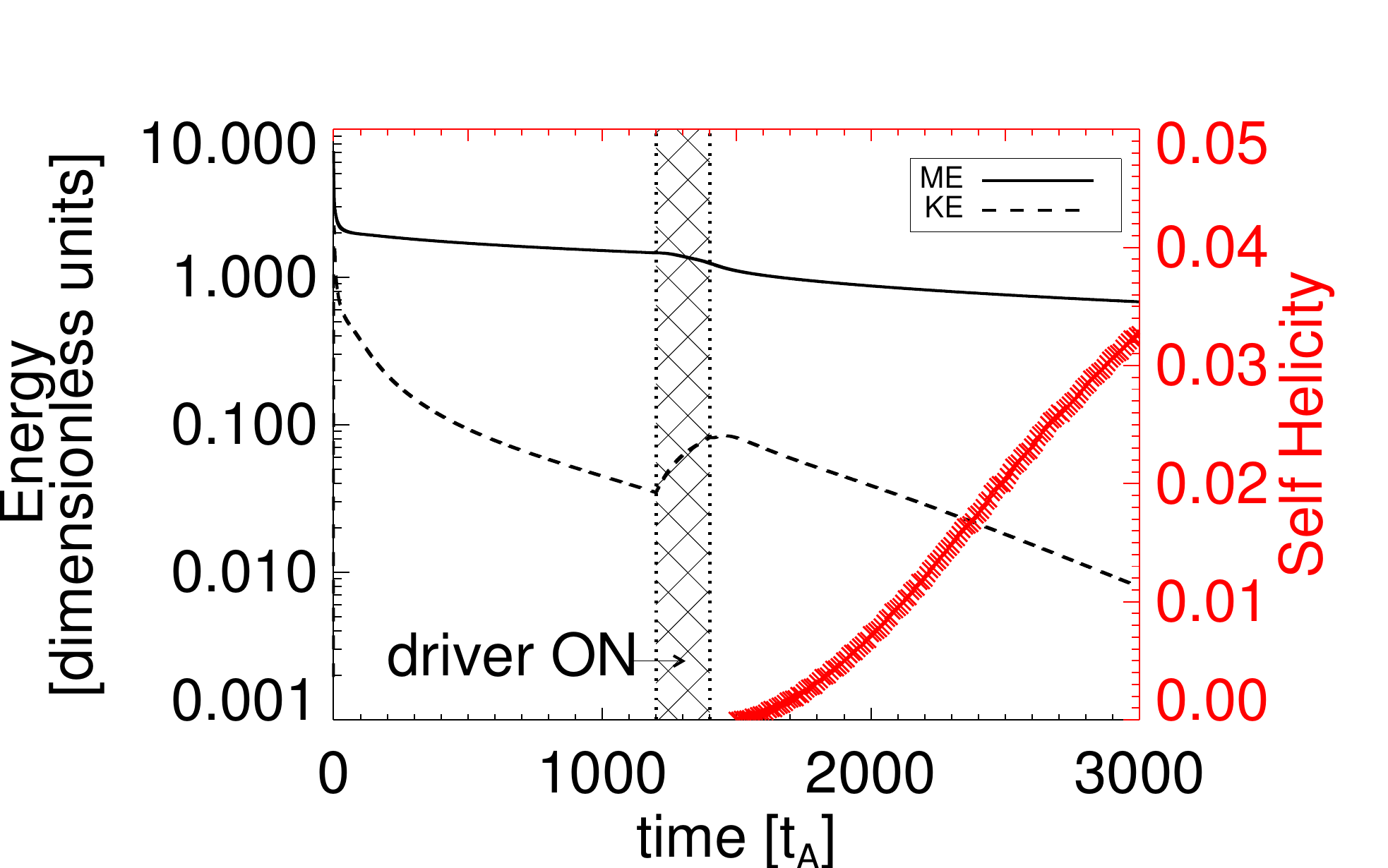}}}
  \caption{{\protect\subref{subfig:Avetwist} Time evolution of the average twist of flux rope field lines calculated in two ways: i) comparing angles of rotation of field line footpoints with respect to the flux rope axis footpoint (red) and ii) integrating parallel current density along each field line (black triangles). \protect\subref{subfig:SelfHelicity} The evolution of self helicity of the system over time (red), together with kinetic and magnetic energies of the system (black) included for context. Energy values are given in dimensionless units, and can be converted to real units using the normalisation values of the simulation; for reference one dimensionless energy unit is equivalent to approximately $3.98\times10^{-16}$J or $2.4\,$\unit{keV}.}}
 \label{fig:twistandhelicity}
\end{figure}
{In Figure~\ref{fig:twistandhelicity}, we compare the average twist and self-helicity of the flux rope over time, and include the kinetic and magnetic energies of the system to provide context. The average angle of footpoint rotation seen in Figure~\ref{subfig:Avetwist}, is calculated using the method described earlier; creating a grid of initial positions throughout the midplane of the flux rope in each simulation snapshot, tracing field lines back to their photospheric sources, and evaluating the rotation of each field line at one source to its position at the other source relative to the flux rope axis. In order that we may better compare with other works which measure twist, we have also used the method of \citet{paper:BergerPrior2006}, in which a twist parameter ($t_w$) evaluates the integrated parallel current along a given field line, \textit{i.e.}:
\begin{equation}
t_w=\int{\frac{{\bf{j}}\cdot{\bf{B}}}{|B|^2} {\rm{d}}l}, \label{eq:tw}
\end{equation}
where ${\rm{d}}l$ is an arc length along the field.
From Figure~\ref{subfig:Avetwist}, both measures differ slightly in values recovered, but the overall trends are in agreement. A direct comparison of the two measures and how (in certain cases) these values 
are related is given in Appendix~\ref{app:twist}. The average field line twist of the flux rope has a maximum shortly after the shearing phase ends, after which it falls, tending to one turn (or a footpoint rotation close to $2\pi$) by the end of the experiment. 
This result is, in some sense, to be expected. The flux rope forms almost immediately after the shearing phase ends, before expanding to fill the simulation volume. The formation itself occurs through reconnection at a large current sheet above the photosphere created by the shearing motion. This reconnection imparts a large amount of twist (up to $3\pi$ along some field lines) into the newly formed flux rope. Following the flux rope formation, additional reconnection (of smaller, fragmented sub-critical currents below the flux rope) allows more field lines to be brought into the flux rope itself; this (and the lack of overlying field) causes the flux rope to expand, while much of the newly added field is twisted at or slightly below $2\pi$, reducing the average twist over time.}

{The energies seen in Figure~\ref{subfig:SelfHelicity} also follow this pattern. As expected, our configuration contains much more magnetic energy than kinetic throughout the experiment. The background resistivity causes a slow decrease in magnetic energy over time, but the fastest rate of magnetic energy release (other than during the very early relaxation phase) can be found immediately and briefly after the shearing phase ends. At this point anomalous resistivity has begun to act on current above the critical value. The shear phase also briefly injects a small amount of kinetic energy into the system, but in general the system continues to lose kinetic energy over time. Using the footpoint field line rotation as the measure of twist, Figure~\ref{subfig:SelfHelicity} suggests that the self-helicity of the flux rope steadily increases from zero over time. The increase in self-helicity seen in Figure~\ref{subfig:SelfHelicity} results from increasing amounts of magnetic flux in the tube. As the magnetic flux rope expands, magnetic reconnection below the tube continues to draw in new magnetic flux, contributing to the flux rope expansion and increasing levels of flux, while slowly reducing the average field line twist.}

The mutual helicity can be found by using the angles $\theta_1$ and $\theta_2$ (for flux tubes ${\rm{A}}$ and ${\rm{B}}$ lying parallel to one another) or $\theta_3$ and $\theta_4$ (in the case of a flux tube ${\rm{C}}$ that overlies a second flux tube ${\rm{D}}$), identified in Figure~\ref{fig:zip2}, namely that
\begin{numcases}{H_{\rm{T}}=}
  \frac{\theta_2-\theta_1}{2\pi}F_{\rm{A}} F_{\rm{B}}, & {\rm{(pre-reconnection configuration)}}\\
  -\frac{\theta_3-\theta_4}{2\pi}F_{\rm{C}} F_{\rm{D}}. & {\rm{(post-reconnection configuration)}}
\end{numcases}  
However, the total magnetic flux through a specific contour of $B_z$ at the base is not constant due to the broadening of magnetic flux sources.

Using conservation of helicity, \citet{paper:PriestLongcope2017} estimated that both the overlying and underlying flux tubes would have a twist of $\pi$, provided that the self-helicity is split equally between the two tubes. In our experiment, however, diffusion at the computational base implies that only an overlying tube forms. Conservation of total magnetic helicity implies that its average twist will therefore be $2\pi$, as observed, since all the self-helicity goes into the overlying tube.

{Finally, we will broadly place our findings in the context of other recent investigations. In the event studied in \citet{paper:Wangetal2017}, two reconnection phases are recovered; however, the first phase \citep[which, in the conceptual model of][would correspond to the zipper reconnection phase]{paper:PriestLongcope2017} displays characteristics of both zipper reconnection and main-phase reconnection. This rules out a direct comparison between the temporal evolution of magnetic twist in both cases. Our recovery of a magnetic flux rope with an average twist of 2$\pi$ might initially appear to be much less than the typical values of twist found by \citet{paper:Wangetal2017}. However, we note that our model represents only one single component (a so-called ``zippette") of the full zipper reconnection picture (which consists of many zippettes occuring one after the other). Now that we have shown it is possible to form a magnetic flux rope through this mechanism, the next step would be to investigate how several zippettes might combine to form a full zipper reconnection event, and how each zippette event contributes to the evolution of magnetic twist in such a flux rope. Such a model would provide a better comparison for the observations of \citet{paper:Wangetal2017}. Our twist values are much more closely aligned to those found in \citet{paper:Inoueetal2018}. Indeed, in Figure~\ref{subfig:Avetwist} we evaluate the field line twist using the same method, and recover values which agree well with those found in their model \citep[noting that we only evaluate the twist parameter of field lines which lie within the flux rope, while][calculate this for many field lines, and use this to infer that a twisted flux rope has indeed formed on some of those field lines, when $t_w$ increases above unity]{paper:Inoueetal2018}. However, we would also emphasise that a direct comparison between values of $t_w$ and the twist given by the angle of footpoint rotation relative to axis footpoints, even in relatively simple cases, can be complex and potentially misleading
\citep[as shown in Appendix~\ref{app:twist}, and discussed further in \textit{e.g.}][]{paper:Liuetal2016}.}

\section{Conclusions and Future Work}\label{sec:conc}
In conclusion, we have demonstrated that a helical flux rope can be formed directly by shearing and reconnection of two untwisted flux tubes (without the need for an inflow velocity). We have examined the properties of this system in detail, observing that the outermost flux sources in the system preferentially undergo reconnection. Although the sources overlap and diffuse slightly at the base, we find that the average level of twist in the flux rope tends to $2\pi$ by the end of the experiment. This twist can be explained simply by evoking total magnetic helicity conservation and so equating the final self-helicity of the flux rope to the initial mutual helicity of the system.

This is only the first step in an investigation into the mechanism of zipper or zippette reconnection. Whether and how a sequence of such events can form a single erupting flux rope overlying a magnetic arcade of post-flare-loop-like structures remains to be seen. The magnetic topology present in this configuration remains relatively simple with the initial configuration consisting of two untwisted flux tubes side by side.
In future, we plan to conduct experiments with much more realistic magnetic structures and to decrease the diffusion near the base, so as attempt to show how a single erupting magnetic flux rope overlying a magnetic arcade of flare loops may be produced.

Key questions in future include: Is the same reconnection achieved if the positive and negative sources do not overlap? What is the effect of modifying the form of the velocity driver applied at the base to move the flux sources as solid bodies? What is the effect of reducing the diffusion near the base so as to prevent the sources spreading out?

\begin{acknowledgements}
The authors gratefully acknowledge the financial support of STFC through the Consolidated grant,  ST/N000609/1, to the University of St Andrews. This work used the DIRAC 1, UKMHD Consortium machine at the
University of St Andrews, the DiRAC Data Centric system at Durham University, operated by the Institute for Computational Cosmology, and the DiRAC Data Analytic system at the University of Cambridge, 
operated by the University of Cambridge High Performance Computing Service. These systems are operated on behalf of the
STFC DiRAC HPC Facility (www.dirac.ac.uk). The equipment was funded by BIS National
E-infrastructure capital grants (ST/K00042X/1 and ST/K001590/1), STFC capital grants (ST/K00087X/1, ST/H008861/1 and ST/H00887X/1) and
DiRAC Operations grants (ST/K003267/1 and ST/K00333X/1). DiRAC is part of the National E-Infrastructure.
The research data supporting this publication can be accessed at \url{http://dx.doi.org/10.17630/dea50f38-7aff-48c3-b090-c689d02dd305}
\end{acknowledgements}

\section*{Disclosure of Potential Conflicts of Interests}
The authors declare that they have no conflicts of interest.

{\appendix\section{Comparison of Twist Measures}\label{app:twist}
Throughout this paper, we have used the ``twist'' to mean the angle a field line rotates through in going from one footpoint to the other, relative to the axis of the loop. 
This is the definition used by \citet{paper:HoodPriest1979} when analysing the kink instability. 
{The link between this definition of twist and the integral expression $t_{w}$ (which is the value of the force-free parameter, $\alpha$, times the length of the field line) is often unclear \citep[see the detailed theoretical comparison in Appendix C of][and the discussion therein]{paper:Liuetal2016}. We will practically demonstrate how these two quantities are linked,}
in the simple case of a straight cylindrical, force-free loop, which allows the two expressions to be evaluated analytically and directly compared.

The angle a field line rotates through is simply
\begin{equation}
\Phi (r) = \frac{L B_\theta}{r B_z}\; ,
\end{equation}
where $L$ is the length of the loop and $B_\theta$ and $B_z$ are the azimuthal and axial magnetic field components. The axis of the loop is at $r=0$. 
$\Phi(r)$ can be prescribed, giving the twist as a function of radius, $r$. 
For example, \citet{paper:Hoodetal2016} used $\Phi(r) = \Phi_0 (1 - r^2/a^2)^3$, for a loop of radius $a$. If the field is in force-free equilibrium, then
\[
\frac{B_\theta}{r}\frac{{\rm{d}}}{{\rm{d}}r} (r B_\theta) + B_z \frac{{\rm{d}}}{{\rm{d}}r} (B_z)= 0\; .
\]
Eliminating $B_\theta$, we have force balance provided $B_z$ satisfies the equation
\begin{equation}
\left ( 1 + \frac{r^2 \Phi^2}{L^2}\right ) \frac{{\rm{d}}B_z}{{\rm{d}}r} + \frac{\Phi}{L^2}\frac{{\rm{d}}}{{\rm{d}}r}\left (r^2 \Phi \right ) B_z = 0\; .
\end{equation}
If the twist is constant, $\Phi = \Phi_0$ and $B_z = B_0/(1 + r^2/a^2)$. 

Turning to the integral expression for $t_w$ (given in Equation~\ref{eq:tw}), using the expressions for $B_\theta$ in terms of $B_z$ and $\Phi$ and taking ${\rm{d}}l = (B/B_z) {\rm{d}}z$, we have
\begin{equation}
t_w = \int_{z=0}^L \frac{{\bf j}\cdot {\bf B}}{B^2} \frac{B}{B_z} {\rm{d}}z = \frac{1}{r}\frac{(r^2 \Phi)^\prime}{\sqrt{1 + r^2 \Phi^2/L^2}}\; , \label{eq:tw2}
\end{equation}
where $\prime$ denotes a derivative with respect to $r$.
Thus the link between the twist angle that we use and $t_w$ is complicated and depends on the radius. For the simplified case of a constant twist field, Equation~\ref{eq:tw2} reduces to
\begin{equation}
t_{w0} = \frac{2 \Phi_0}{\sqrt{1 + r^2 \Phi_0^2/L^2}}\; .
\end{equation}
Figure \ref{fig:app1} shows the variation of $\Phi$ and $t_w$ with $r$ for the twist profile defined above.
\begin{figure}[t]
\centering
 \resizebox{0.99\textwidth}{!}{\includegraphics[clip=true, trim=20 10 10 10]{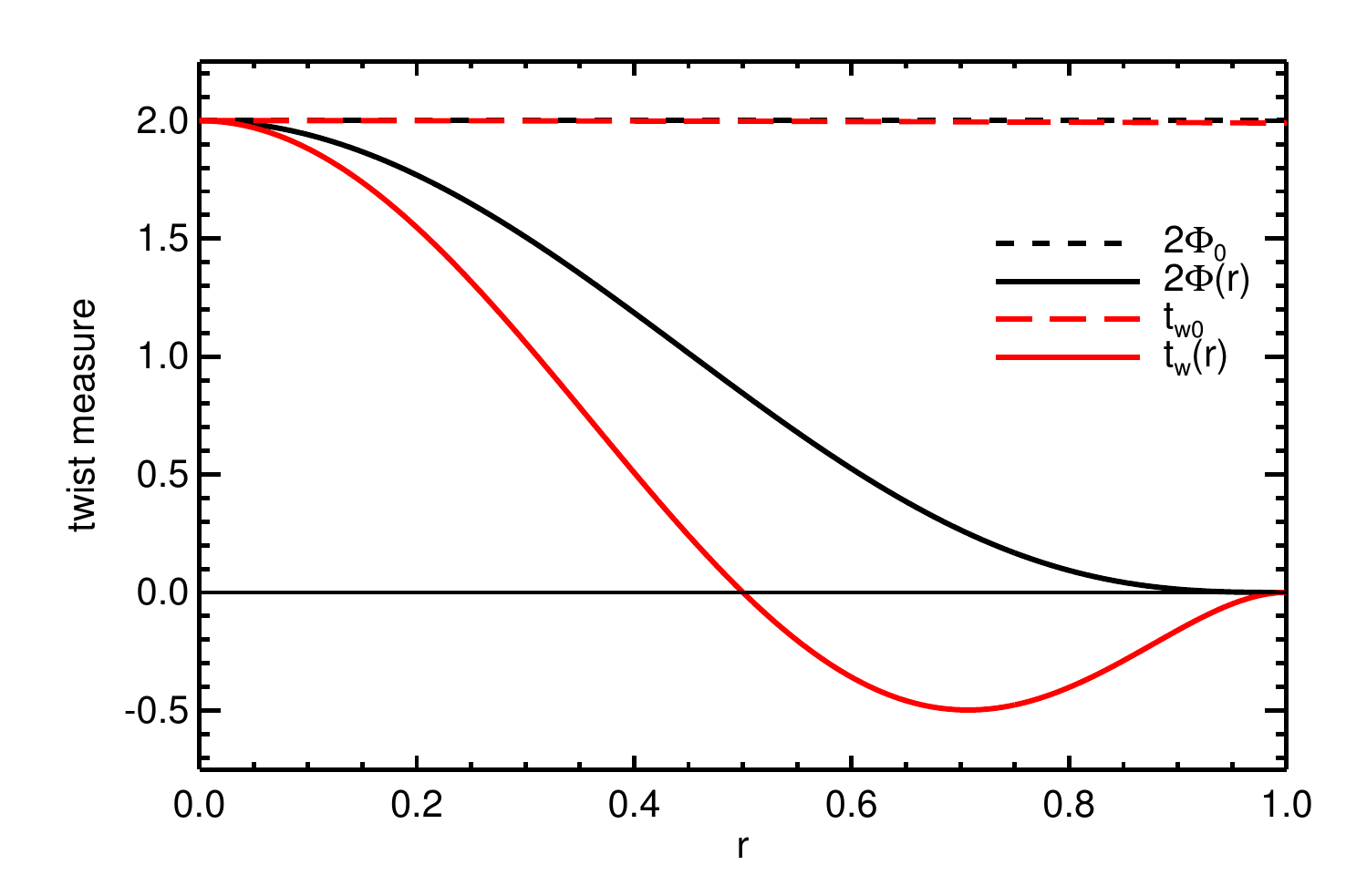}}
  \caption{Radial variation of footpoint twist ($\Phi$) and integrated parallel current ($t_w$), using example twist profile defined by \citet{paper:Hoodetal2016}, where $L=10$, $a=1$ and $\Phi_0=1$. Uniform twist profiles are seen as dashed lines, while radially varying profiles are thick solid curves (for key, see legend).}
 \label{fig:app1}
\end{figure}

It is clear that $2\Phi_0$ and $t_w$ are almost identical for the constant twist field. However, there is a significant difference between the two measures when the non-uniform twist profile is used. In particular, while $\Phi$ remains positive throughout the loop cross section, $t_w$ changes sign. In addition, if we calculate the average value of $t_w$ across the circular cross section of the loop for which $r \Phi \ll L$, then the value is zero. This is not true for the twist.

While it is computationally easier to calculate $t_w$ in a more general, non-cylindrical loop, we believe {that (where possible)} it is better and clearer to define the twist as the angle a field line rotates about the loop axis.}

\bibliographystyle{spr-mp-sola}
\bibliography{SOLA-D-18-00039}  

\end{article} 

\end{document}